# Modeling Context, Collaboration, and Civilization in End-User Informatics

## George A. Maney

*Abstract*

End-user informatics applications are Internet data web management automation solutions. These are mass modeling and mass management collaborative communal consensus solutions. They are made and maintained by managerial, professional, technical and specialist end-users.

In end-user informatics the end-users are always right. So it becomes necessary for information technology professionals to understand information and informatics from the end-user perspective.

End-user informatics starts with the observation that practical prose is a mass consensus communal modeling technology. This high technology is the mechanistic modeling medium we all use every day in all of our practical pursuits.

Practical information flows are the lifeblood of modern capitalist communities. But what exactly is practical information? It's ultimately physical information, but the physics is highly emergent rather than elementary. So practical reality is just physical reality in deep disguise.

Practical prose is the medium that we all use to model the everyday and elite mechanics of practical reality. So this is the medium that end-user informatics must automate and animate.

Table of Contents

# 1: The General Theory of Practical Information

## *Introduction*

What is practical information? What does it mean? How does it work? These are the questions answered by The General Theory of Practical Information (GTPI). This theory answers these key questions in a very particular and perhaps even peculiar way.

This theory provides a ruthlessly, relentlessly mechanistic perspective for understanding practical models, meaning, and management. It draws heavily on classic models from the mechanistic foundations of the hard sciences and the high technologies.

This firm focus on mechanism starts with three observations:

- we live in a mechanistic universe which we all must cope with every day
- the meaning of practical information is primarily mechanical
- computers are merely mechanistic modeling machines

These observations are the starting point for the mechanistic sort of machine intelligence that we've been using for nearly a hundred years now. This is the sort we started with in the days of punch cards. This is the sort that makes personal computers work. This is probably just the same sort of machine intelligence we will be using a hundred years, or even a thousand years, from now.

Thus for the foreseeable future practical information in practical informatics applications must be understood in terms of mechanistic models and mechanistic meaning. Today we're not quite sure where the limitations of mechanistic modeling ultimately lie.

Perhaps the meaning of everything in physical and practical reality will eventually be explained at some level of emergent mechanics. Perhaps we will never be able to explain much more with mechanistic modeling than we do today.

Is Shakespeare practical? The body of Shakespeare's work may well be the single greatest source of secular practical wisdom ever assembled. Even so, this sort of practical meaning is not readily reducible to mechanism by any means available today.

Here's a pop quiz. Turn your favorite Shakespearean sonnet into an algorithm. For extra credit, turn your favorite algorithm into an English sonnet.

Why are these challenges absurd? The answer is that we have no idea how automate the meaning of poetry today. All applicable algorithms automate and animate applied mechanics and applied mathematics models.

If The General Theory of Practical Information can't explain Shakespearean information then what's it good for? It's good for explaining applied mathematics and applied mechanics information. It's good for explaining basic, natural, and applied science information. It's good for explaining enterprise, economics, and engineering information. It's good for explaining the mechanistic information of all everyday and elite practical pursuits.

## *Practical Prose as High Technology*

The only information technology ever employed in any practical pursuit is the one that you yourself are using right this very instant. This technology is practical prose. All other kinds of practical information technology are derivative of practical prose.

Practical prose is that separable and specialized form of prose that we all use to model the dynamics of physical reality and practical reality. There are, of course other kinds of prose. Some examples include poetic, personal, philosophical, pious, and profane prose.

So far we really don't understand much about the meaning of these other kinds of prose. The mechanistic meaning of practical prose is the only sort of prose meaning we have any systematic understanding of today.

Practical prose is the most important technology in your life. This is the technology that you use to communicate and collaborate with others in all of your practical pursuits. It is a technology that you have spent many thousands of hours mastering.

The entire practical dynamics of everything in human civilization is modeled by means of practical prose. For this reason practical prose is the indispensable universal ingredient in everything we use and everything we do.

Practical prose is the mechanistic modeling medium of every practical pursuit in all of human civilization. So practical prose is necessarily the starting point for every other technology in human civilization.

In fact practical prose is the one truly indispensable technology in any human civilization. This is a point made in antiquity by the story of Babel in Genesis 11.

Practical prose is the ultimate capitalist tool. Practical information flows are the lifeblood of modern capitalist communities. Practical prose is the medium of these practical information flows.

Practical prose is the product of ten thousand years of technological evolution. It is a mass consensus communal technology. It is an organic technology. It is an intensively self-ordering and self-optimizing mass technology.

For this reason practical prose is perfect for all practical purposes. It is perfect in much the same way, and for much the same reasons, as free financial markets.

The existence proof for the perfection of practical prose is all around us. Practical reality itself is the sum and substance of this proof. Everything that is made or moved or managed in practical reality must be adequately modeled by means of practical prose. Anything else is miraculous. There are no accidental artifacts.

At this very moment millions of people are working to maintain the perfection of practical prose. Today this mass social process continues around the world around the clock on the Internet.

Remarkably, practical prose has long been mostly an orphan. The last serious and systematic effort to understand practical prose technostructure took place in the 1920s and 1930s.

Yet practical prose has a remarkably elegant and elaborate technostructure. Anyone with good practical prose skills can start uncovering large chunks of this technostructure after a few hours of coaching. Exotic sensibilities and esoteric skills are not required.

## *Practical Information*

Practical information flows are the lifeblood of modern capitalist communities. But what exactly is practical information?

For present purposes practical information is any mechanistic model of practical reality. Practical information models are expressed in terms of practical prose. Practical prose models always consist of two closely correlated parts:

- practical logic: mechanistic or mathematical exposition of practical reality meaning
- practical language: verbal or visual encodings of practical prose logic

There are no unmodeled or unmanaged artifacts in the modern industrial world. So every artifact is necessarily supported over its entire life by suitable practical prose models. This includes complimentary mundane and managerial models.

The meaning of these models starts with practical reality. Practical reality is the reality of everyday human experience. This is the familiar reality of people, places, periods, property, practice, policy and such.

Practical reality is an emergent view of underlying physical reality. Physical reality is the foreign reality of molecules, momentum, minerals, and such.

The basics of physical and practical reality are summarized as:

| Reality Perspective | Spacetime | Substance | State |
|---|---|---|---|
| physical reality | space & time | matter & energy | phenomena & proportionality |
| practical reality | period & place | person & property | principle & practice & policy |

The key idea here is physical state. This is a basic idea from your secondary school physics textbook. Physical state is the elementary state of physical reality. Practical state is the emergent state of practical reality.

Are you a stateless person? Certainly not. You've got lots and lots of kinds of physical and practical state.

Your physician is concerned with the state of your health. You broker is concerned with the state of you finances. You boss is concerned with the state of your work assignments.

We exist in physical reality but don't generally experience this reality. Emergent practical reality is the reality of everyday living in the modern world. The everyday mechanics of this reality are about making, moving, maintaining, managing, modeling, and so on. These are the mechanisms we all must master in order to cope with real life in the real world.

The ultimate reference model in every practical pursuit is present practical reality. This is the entirety of the world of practical pursuits as it is at this very moment.

Present practical reality is the starting point for every valid practical information model. The symbolic information models employed in all practical pursuits are derivative of this ultimate practical information model.

Thus you yourself are an information model. You are a substantial information model rather than a symbolic one. Your driving license is a symbolic model of you.

You are the best possible present model of you. You are, to some extent, a model of everyone else. This is true because you are similar, in some ways, to everybody else. Practical significance starts with practical similarity.

In this sense information is imitation. Representation reflects and resembles reality. Models are the mechanisms we use to mirror and mimic other mechanisms.

Everything in physical and practical reality is a substantial self-information model. In all practical pursuits symbolic separate-information models such as drawings and documents are reflections of physical and practical reality.

We live in a mechanistic universe. So the entire practical meaning of both substantial and symbolic information models in practical pursuits is necessarily mechanistic. All other sorts of meaning are moot for all practical purposes.

There is much about the mechanics of our universe that we don't understand. Even so, there is an awful lot that we do understand. The part that we do understand so far is the starting point for meaning, modeling, and management in all practical pursuits.

Human civilization is the ultimate communal system. Communal systems operate by means of communication and collaboration. Practical prose is the medium of practical communication and collaboration in human civilization.

Human culture is the ultimate information system. The culture of any community is the totality of every substantial and symbolic model available in the community. Culture is the basis for communal communication and collaboration in every human community.

Practical culture is only one part of human culture. Even so, it's a particularly important part. This is the part we use to cope with living out our lives in present practical reality.

## Practical Information Ideologies

Call up you favorite IT expert sometime and ask for quick explanation of practical information. In all likelihood you will find the resulting explanation utterly incomprehensible. Even so, this is not a problem. Don't panic. This is exactly as it should be.

The reason is that there are two primary ideologies of practical information today. These include:

| Information Modeling Ideology | Prevalent in | Practical Information is | Foundations |
|---|---|---|---|
| universal modeling | programming | computation in disguise | metaphysical |
| utilitarian modeling | practical pursuits | civilization in disguise | mechanical |

These two practical information ideologies could hardly be more different. Even so, both ideologies work exceptionally well. The one that you actually need all depends on who you are and what you're doing.

Most software professionals see practical information as computation in disguise. This is the starting point for universal modeling in best practice information technology. This includes all commonplace forms of data engineering and knowledge engineering[1].

The best software authors cultivate specialized skills that allow them to see the dynamics of practical reality directly in terms of computational algebras and algorithms. This is demonstrably the optimal way of looking at practical information in current best-practice software production.

Yet this is just one way, not the only way, of looking at practical reality. Moreover, this is a way of looking at practical reality that is utterly foreign to practical informatics end-users.

Practical informatics end-users just don't understand computational algebras and algorithms. They never have and they never will. Nor should they.

These practical folks have an entirely different way, a far more powerful and precise way, of looking at practical reality and practical information. They see practical reality as a great diversity of mundane and managerial mechanisms.

In this way practical informatics end-users see practical information as civilization in disguise. This is the starting point for utilitarian modeling in all principled practical pursuits.

The dichotomy of information ideologies comes at a steep price. It is this dichotomy that explains why software professionals have such great difficulty in understanding what their end-users really want and need. It also explains exactly why practical informatics software, despite often heroic efforts on the part of software professionals, is so remarkably difficult to use.

What is a computer? The answer depends on your information ideology. Programmers see computers as marvelous mathematical machines. Informatics end-users see computers as machine tools of mass modeling and mass management.

Computers are not in any sense magical. Computers, for all practical purposes, are just another sort of machine. These are remarkably complex machines but just machines nonetheless.

In practical pursuits computers are practical information machines. They are the machines that we use to mirror and mimic the micromodular mechanics of practical reality.

## Practical Management

Mankind is managerial. This is hardly the only way of understanding mankind. Even so, this is always the best mechanistic way to understand mankind for all practical purposes.

Managing is mandatory for all competent adults in the modern industrial world. We must manage our own affairs and often those of others. Some very few of us manage the elite affairs of enormous enterprises. Yet all of us are necessarily expert everyday managers.

At the very least we usually manage to get out of bed in the morning, go about our business, and crawl back into bed at night. So all of us are always managing something. Mostly we are managing many things.

You may well be skeptical. If so, then try to go for an entire day without managing anything. You will find that this is simply impossible unless you have been bound, gagged, and tied to a chair.

This is exactly why practical information is so vitally important in all practical pursuits. Practical information is always the fruit of practical management as well as the feedstock. So practical models are the stock in trade of practical management.

The foundations and fundamentals of practical management are essentially the same in all practical pursuits. They are just the same in managing daily living as they are in managing the dealings of multinational corporations.

The primary varieties of management models are measurement and mandate models. Measurement models capture the meaning of the working world as it is. Mandate models capture the meaning of the working world as we wish, want, or will it to become.

Managerial information is mostly messy and multitopical. This is particularly true of general management information. General management information is often massively multitopical.

The quality of practical management can never be better than the quality of the practical prose models available for managing. This is true both for enterprise management and for everyday management.

Why must we manage? The answer is that all of the lifeforms we know of engage in management. Dogs somehow manage to be dogs. In this sense every lifeform is a stable, self-managing mortal mechanism.

Nearly all lifeforms engage in both self and separate management. Predation and parasitism are examples of separate management in the natural world.

Management, in this view, is any form of complex conative control. Every species has its own purposes and practices. So there are endless specialized forms of management in the natural world.

You yourself are full of these natural management mechanisms. These are the diverse homeostatic control mechanisms that manage the physiological balance of your body.

Note that genetic information is management information. A genome is the natural information system employed to manage the evolution of a species. The genome of a species encodes all of the anatomical and physiological mechanism possibilities of that species.

Human management is unique. Human beings have the capacity to manage collaboratively based on the communication of shared mechanistic models. Other species, such as honeybees, do this as well. The difference is that our management models are complex and cultural rather than circumscribed and congenital.

We manage by means of principles rather than patterns. We manage by means of intellect rather than instinct. In these ways we can and do master and manage everything within the range of available phenomena, principle, and practice mechanism models.

These models are practical prose models. This is why understanding practical prose is uniquely important in understanding *Homo sapiens* and human society. Every species on this planet has a genome. Only human beings have a practical prosome.

All human civilizations are communal management systems. Modern mass civilization in the industrial world increasingly relies on mass management and mass modeling.

These are the means by which we manage mass mundane systems. These are mass making and moving systems. They include mass production, mass distribution, and mass consumption systems. They include mass communication, mass transportation, and mass culture systems.

Today we are bred, born, and buried by means of mass systems. We are educated, employed, and entertained by means of mass systems. This is a radically different way of living in the world that that which prevailed anywhere only a century ago.

Modern mechanistic management makes this new way of living possible. We live in the first methodically, meticulously, minutely modeled and managed human society. All this modeling and managing is the genesis of our unprecedented material wellbeing and monetary wealth.

## *Practical Prose Technostructure*

What kind of technology is practical prose? It can't be a mundane technology like a dishwasher or an automobile. Practical prose doesn't make anything or move anything.

That's because practical prose is a managerial technology. It's the technology of practical management measuring and mandating. It's the technology of mechanistic practice modeling.

In what sense is practical prose a high technology? In the sense that, as in the case of all high technologies, we understand much of the high mechanics. We understand multiple layers and levels of mechanics from the elementary to the emergent.

The technostructure of practical prose is hidden in plain sight. This is why we usually don't usually pay any attention to this truly marvelous technological wonder.

This is to say that we don't usually intellectualize practical prose mechanistic meaning. The reason for this is that we all spend many years systematically internalizing this meaning.

We internalize this meaning over time as we learn to work in the world. Play is the work of children. We begin to explore the mechanics of practical reality at a tender age. Toddlers tinker endlessly with everything. These are our early efforts at empirical experimentation.

This sort of play is the beginning of our mastery of mechanistic meaning. This play is how we lay down the foundations of our working worldviews.

We build on these foundations gradually over time. We garner education and experience. We encounter and explore diverse practical reality micromechanisms. We wire the language and logic of these into our working worldviews incrementally. Every now and then we remodel parts of our working worldviews by wiring them up in new and better ways.

We continue to cultivate our working worldviews throughout our lives. This is how practical language and logic become second nature to us.

Since practical prose is second nature we must develop specialized skills in order to explore its technostructure. These skills involve intellectualizing the mechanistic meaning that we have spent years laboriously learning to internalize. These skills involve visualizing the micromodular structure of our own personal working worldviews.

Most people can learn to explore practical prose technostructure with a little practice. Some people seem to be much better explorers than others. Some find this sort of exploration inordinately difficult.

## *Practical Prose Exposition Frameworks*

Practical reality is a massive, multifarious, messy place. Its complexity confounds us. Its diversity daunts us. Its mechanisms are far more random than regular. How do we manage to model all this madcap mechanistic messiness?

The answer is that we divvy everything up. We divide and conquer. We map the multifarious mechanisms of practical reality in to a great diversity of empirical exposition frameworks.

Some practical exposition elements are atomic but most are not. Most exposition elements are included in a major or minor exposition framework.

These frameworks are constructive spaces for modeling particular sorts of mechanism like time or terrain. From a linguistic science semantics standpoint this is somewhat similar to prototype theory[2].

We map out exposition frameworks as constructive spaces of practical exposition elements. These practical exposition frameworks are of three primary types:

- rationale: artifact application frameworks
- rhetoric: artifact applied mechanics frameworks
- rigor: artifact applied mathematics frameworks

Each framework provides a unique rationale, rhetoric, or rigor modeling perspective. These perspectives serve to constructively cover a range of practical meaning. Some examples of expository framework perspectives are:

- rationale perspective example: all of the various breeds of show dogs
- rhetoric perspective example: all of the Julian calendar periods
- rigor perspective example: all of the usual integer arithmetic operations

Practical expression often involves all three primary types of exposition elements. Consider a trivial example:

- rationale element: apple
- rhetoric element: collection
- rigor element: count

Counting a collection of apples is commonplace. Counting a collection of oranges is much the same thing. If your collection of apples has different kinds of apples then a count of classes might be useful.

Practical prose typically draws on many exposition element frameworks simultaneously. We often use dozens or many dozens in simple practical prose models. This combination of practice frameworks perspectives provides:

- redundancy: lots of alternative elements
- reusability: lots of ways to use each elements
- recombinance: lots of ways to put the elements together

Putting these framework perspectives together involves:

- pluralism: multiple kinds of complementary perspectives
- parallelism: mutual correlation of complementary perspectives

It is exactly this pluralism and parallelism that give practical prose its extreme range of representational power, precision, and profundity. Small reductions in available pluralism and parallelism result in large reductions in available representation quality.

High quality exposition model structures are characterized by:

- referential richness: lots of referential meaning
- randomness: mostly random with local regularity
- reinforcement: multiple parallel perspective mashups

All of this is exactly what we would expect. The reason is that practical exposition mirrors and mimics the practice dynamics structure of practical reality. Practical reality is characterized by:

- extreme diversity of form and function
- lots of micromechanical similarity
- less macromechanical similarity

- large scale disorder emerging from diverse local order
- constant dynamic coevolution of changing combinations

The wrong way to build an exposition framework is to enumerate practical reality. Perhaps the gods do this. Perhaps the gods have a name for every individual grain of sand.

We mortals, on the other hand, can hardly manage this. So we simplify practical reality by finding as much similarity as possible. This includes both similarity of form and similarity of function.

We capture this similarity in terms of applied mechanics and applied mathematical micromodels. These similarity frameworks act as a meaning compression mechanism. They serve to filter the redundancy of real life meaning out of all the randomness. The electromagnetic equations, the balance sheet equations, and the diurnal cycle are examples of this sort of filtering.

Mechanistic meaning similarity is particularly important in the applied mechanical rhetoric of practical prose. Here we see a vast diversity of:

- metamechanisms: reusable micromechanism prototypes
- mechaphors: similarity structures among mechanisms

Everyday practical prose context frameworks are highly efficient systematic compression mechanisms. Without these principled practice frameworks we simply couldn't cope.

These everyday coping frameworks are hardly simple. That's why it takes us all about two decades to master these frameworks. We begin with mundane coping frameworks and progress to managerial coping frameworks.

Isn't there some secret simplicity that underlies all this sophistication? Apparently not. If there is then we haven't seen a trace of it so far.

Moreover, there is good reason to believe, based on what we know today, that this secret simplicity cannot exist. Combining a few simple physical laws creates phenomena spaces of staggering complexity and extreme sparseness.

This is the kind of complex order that we experience in our daily lives. This is the kind of complex order that we must deal with in our everyday practical pursuits. This is the kind of complex meaning we all model every day with practical prose exposition framework constructions.

## *Practical Language*

Practical language is the medium of mechanistic modeling. Varieties of practical prose can be characterized primarily in terms of:

- visual or verbal style
- qualitative or quantitative substance

Visual styles are the sort we use most often for highly complex modeling. These are convenient when words fail us. Examples include roadmaps, wiring schematics, and PERT charts.

Some simple quantitative styles of practical language include numbers, numeric data sets, and numeric equations. More sophisticated styles include algebraic geometry as well as applied analytic mathematics equations partial differential equations such as Poisson's equation.

Isn't human language a natural phenomenon? The answer is yes and no. The style of spoken practical language is natural. The substance of all forms of practical language is normative.

This stands entirely to reason. We are not born with innate models of the mechanics of baking a cake or driving a car. We acquire these technological models over time by means of education and experience.

Practical prose models are empirical expressions. The elements of empirical expression are:

- encoding: the variable linguistic style of expression
- exposition: the invariant logical substance of expression

Consider the operating manual for an automobile. In North America today these usually combine English, French, and Spanish translations. These are three very different forms of encoding. Even so, the exposition is essentially the same in each translation.

The range of practical language includes the entirely of:

- applied mechanics models and modeling elements
- applied mathematics models and modeling elements

The applied mechanics and applied mathematics of practical prose come from three basic kinds of practical rationales. These include:

| Principled Technologies | Phenomena Theory | Practical Dynamics |
|---|---|---|
| enterprise principles | production phenomena | work dynamics |
| economic principles | price phenomena | wealth dynamics |
| engineering principles | potential phenomena | world dynamics |

In most routine practical prose models enterprise principles are the dominant rationale. Production and praxeology (systematic production management) primitives are typically the starting point for routine practice modeling.

## *Practical Logic*

There are lots of different sorts of logic. At some point in your education you may have struggled thought a course on predicate logic. This is the sort of logic with all the reversed Es and inverted As. This is a commonplace form of mathematical logic.

Mechanical logics are something else entirely. Mechanical logics are empirical logics of physical and practical exposition. Mathematical logics, in practical pursuits, are logics of encoding.

Exposition logics and encoding logics are always separable issues in practical information. You always need both kinds of logic. These are always completely complementary.

In high school physics we are taught Newton's second law of motion:

$F = M \times A$ where:

- F is force
- M is mass
- A is acceleration

Sometimes it's easy to forget that there are two entirely separable logics here. These include:

- the explicit mathematical encoding logic expressed as an algebraic equation
- the implicit mechanical exposition expressed in terms of physical state phenomena proportionality and parameterization

We tend to systematically confuse these two very different kinds of logic, especially in educational settings. Thus it becomes easy to think of mathematics as having practical meaning in and of itself.

Yet in all practical applications mathematics is always a tool and never a truth. Mathematical logics must always be combined with mechanical logics in order to express applied mathematical meaning.

Mechanical logics are the native logics of material reality. They are a consequence of the phenomena structure of physical reality and practical reality. They reflect basic constraints on existence and evolution built into the fabric of the universe.

These constraints include conservation of substance, least action, and the laws of thermodynamics. They also include basic topological and geometric constraints like connectedness, continuity, and compactness.

In practical terms these constraints mean that you can't be at two places at once. They mean that anvils don't appear magically out of thin air as they do in cartoons. They mean that you can't un-fry an egg. They mean that everything always changes.

The pure mechanical logics of physical and practical reality are differential state dynamics logics. These are the pure mechanistic logics of practical language exposition. These are sometimes known as Newtonian logics.

The starting point for these practical logics is the question "What's the difference?" It turns out that this is the only meaningful question we can ask about physical reality and practical reality.

Differentiation is the basis for everything we know about life in our mechanistic universe. We differentiate here from there, then from now, this from that.

The universe just is what it is. Our working models of the universe are all of the diverse differential structures we observe. Our working worldview is a common consensus concerning how to name these differences and how to put them good use in practical pursuits.

Differentiation is necessary and sufficient because the phenomena structure of physical reality is differential. The constraints of natural law combinations dictate that some kinds of differential state dynamics are permitted and that other kinds are not.

Physical state is the starting point for modeling physically feasible differences. Recall from your high school physics textbook that physical state is the dynamic embedding of substance in spacetime.

Practical state, such as the state of your finances, is ultimately just highly emergent physical state. Understanding the deep mechanics of this emergence is not always necessary for practical purposes.

Consider mixing a milkshake. A milkshake is a colloidal suspension with lots of messy emergent mechanics. We don't yet understand much about the turbulent fluid dynamics of these complex chemical suspensions. Even so, we mix an awful lot of milkshakes.

All of us in the modern industrial world are masters of many hundreds of specialized practice logics. Some of us manage to master many thousands.

We start to build our mastery as toddlers. We begin with light switches and doorknobs. We graduate to cable TV remote controls and video games.

How many kinds of practical state dynamics logics are there? There a millions of millions. There are state dynamics logics for telephones and traffic laws. There are state dynamics logics for bookkeeping and bowling. There are state dynamics logics for cooking and chemistry. There are state dynamics logics for airframes and atomic physics.

A road map is a commonplace example of a differential state dynamics logic framework. Roadmaps are constructive collections of spatial differences. These spatial differences are paths between places. Possible routes among places are constructed by combining paths into pathways. Little differences are concatenated to form larger differences.

Consider the example of a roadmap of the U.S. Interstate Highway System. This is a commonplace example of a visual prose constructive practice framework. This framework specifies the state dynamics spectrum of all possible interstate highway trips.

Let's say that you want to drive from Los Angles to Chicago. Many interstate highway pathways can be constructed between these two places, including myriad circuitous and chaotic routes. Choosing the best route from all these candidates is a practical management challenge.

Choosing your route involves lots of different kinds of practical considerations. Some examples include weather, road construction, local traffic conditions, and regional fuel prices.

So the interstate highway map is just one of a plurality of practice model involved in choosing your route. All these different kinds of practical logic must be cogently combined in order to choose your route wisely.

## *Practical Infodynamics and Practical Informatics*

Practical information is useless unless it flows. Nothing happens in practical reality without practical information flows. Practical information flows are the basis of all practical communication and practical collaboration.

Automating practical information involves two kinds of applied information science:

- practical infodynamics: how does practical information flow in practical pursuits?
- practical informatics: how can we advantageously automate these infodynamics flows?

Practical information systems have been around since the dawn of agricultural civilization. In the modern industrial world nearly all practical information systems take the form of mass modeling and mass management systems.

Mass modeling and mass management systems are the bedrock of modern capitalist civilizations. These modern information systems provide infodynamics flows of staggering volume, variety, and value. Thus practical information today, in comparison to every historical standard, is cheap, convenient, copious, and comprehensive.

From the practical infodynamics viewpoint modern industrial civilization is a vast system of mass systems. Most everything done in modern industrial civilization involves one or more mass systems. It is these mass systems that enable and engender the remarkable comfort and contentment of life in the modern industrial world.

Those of us that live in modern mass society tend to take our unprecedented wealth and well-being for granted. It is all too easy for us to forget that billions of our fellow human beings around the world today still live precariously on about a dollar per person per day.

Why are we so rich while they are so poor? U.S. annual per capita GDP is over forty thousand dollars. This figure is about six times higher than that of China and ten times higher than that of India. It is about twenty times higher than Madagascar.

The difference, quite simply, is management. The managerial institutions and infrastructure of modern mass industrial society are the basis of our unprecedented wealth and well-being.

Mass institutions and mass infrastructure require mass information flows. Today only a tiny fraction of these flows are automated by informatics means. Ultimately practical informatics will automate and animate the entirety of mass infodynamics flows.

Automating and animating these flows is the sum and substance of informatics end-user value. This is exactly what informatics end-users have always wanted. It is all that they ever will want. This is the general theory of end-user informatics value.

So getting all these infodynamics flows automated and animated is the ultimate goal of informatics. This can only be accomplished with end-user informatics.

## *Practical Intelligence*

Can we really understand practical prose without first thoroughly understanding the mental mysteries of human cognition? Don't we need a far better understanding of human communities and the human condition? Don't we fist need to completely understand human intelligence, intellect, intuition, and insight?

In fact none of this is necessary. The starting point for understanding why is the following critically important distinction:

| Intelligence Variety | Functionality | Science | Medium | Models |
|---|---|---|---|---|
| individual intelligence | cogitative | psychology | psychic mysteries | mentalist |
| institutional intelligence | collaborative | praxeology | practical prose | mechanistic |

The interesting thing about the term "intelligence" is that it has two very different yet related meanings. In one sense the term "intelligence" refers to individual intelligence as in the psychometric notion of an intelligence quota (IQ). In the other sense the term "intelligence" refers to institutional intelligence as in the case of the Central Intelligence Agency (CIA).

We all exhibit individual intelligence. Even so, we apparently cannot observe the workings of our own intelligence. For this reason the mental models of psychology are mostly still a mystery today. So far we are unable to specify and simulate the dynamics of nontrivial psychological models.

Institutional intelligence is another matter entirely. Institutional intelligence arises whenever two or more people are involved in the course of any sort of practical pursuit.

All practical informatics systems in service today automate and animate institutional intelligence models. This is the only sort of machine intelligence we have needed so far. This is the only sort machine intelligence we will likely need for the foreseeable future.

The entirely of institutional intelligence is conveniently self-documenting by means of practical prose. We can always deconstruct the entire mechanistic meaning of everyday and elite practical prose.

Thus institutional practical intelligence is obvious for all practical purposes while individual intelligence is opaque. We cannot observe the workings of our own individual practical intelligence. We can always completely observe the workings of practical institutional intelligence.

Conveniently this allows us to ignore individual practical intelligence when working with institutional practical intelligence. In fact we can even assume that individual practical intelligence is purely derivative. We can assume that individual practical intelligence is nothing more that myriad, perhaps unique, individual adaptations to the institutional intelligence culture.

The classical notion of individual intelligence has to do with some sort of universal reasoning mechanism hidden somewhere in our brains. In many versions this mechanism is some sort of cranial computer[3].

Yet so far we haven't found any trace of this monolithic mental organ. The likely reason for this is that there just isn't any organ of this sort.

The massively micromodular architecture of the practical prose technostructure suggests a much more likely alternative. In this alternative view individual practical intelligence is mechanistic and massively micromodular in just the same way as institutional practical intelligence.

In this view we have lots and lots of small, specialized reasoning mechanisms in our heads. The specific mechanisms in each set vary widely from person to person. The specific mix of these mechanisms is likely due both to the influence of nature and of nurture. This explains why our individual intelligence is as distinctively unique as our fingerprints.

In this view we cope with the mechanistic complexity of practical reality by recombining and retargeting our stock of mechanistic reasoning mechanisms in real-time. We wire these up into constructions suitable to the mechanics of whatever circumstances confront us in the course of daily living.

## *Practical Superintelligence*

Superintelligent computers have long been a staple of science fiction. HAL-9000 in Stanly Kubrick's classic 1968 film *2001* is a notable example.

Today the notion of radically emergent computational superintelligence is a recurrent theme in the popular science press. Here the notion is that computational superintelligence will evolve and emerge on the Internet over the course of the next few decades. This radical emergence is often termed a "singularity".

A key assumption here is that geometric growth in Internet computing power over time will somehow enable and engender a spontaneous self-ordering of massive software superintelligence. From a purely technical perspective all of this is unlikely, but that's beside the point.

There's no need for science fiction superintelligence. Mass managerial superintelligence has long been a practical reality in myriad practical pursuits.

In fact this sort of institutional superintelligence has been evolving and emerging for at least a century. This evolution was well underway long before the appearance of commercial digital computing equipment.

Here's a simple definition of managerial superintelligence:

- a mass management system involving a diversity of institutions and individuals
- stably self-managing, self-ordering, and self-optimizing practical information flows
- capable of managing high-complexity practice that would otherwise be unmanageable

Managerial superintelligence is any institutional capability to model and manage that substantially exceeds that of individual sapient intelligence. This is the synergistic superintelligence of mass modeling and mass management. It is the combinatorial superintelligence of massive collective managerial competence. It is the emergent superintelligence of managerial mass communication and mass collaboration.

Modern mass society simply could not exist without managerial superintelligence. This mass synergistic intelligence is the starting point for the myriad of massively complex practice systems that make modern mass society work.

Note that managerial superintelligence doesn't require superintelligent managers. In fact quite the opposite is true.

The best sort of managers for managerial superintelligence structures are those with mature specialized management skills and savvy. Good practical communication and collaboration skills are a must. Exceptional individual intelligence is not required and may actually be counterproductive.

How does managerial superintelligence work? Consider your car. In 1900 a single person could master everything then known about automotive products, processes, and plants. Today this is inconceivable. No single person is nearly smart enough to model the entire lifecycle of a modern automobile.

An automotive product lifecycle involves many hundreds of sorts of specialized topical modeling. This lifecycle involves many thousands of sorts of specialized target modeling. All of these models must be adequately and accurately combined in order to manage the entire lifecycle of a mass produced automotive product.

There are product and plant models. There are manufacturing and marketing models. There are sales and service models. There are economic and environmental models. There are legal and logistics models. There are training and tooling models. Then there are many more kinds of models as well.

Legions of managers somehow manage to put all the pieces of this vast mass modeling puzzle together. Successful mass management requires that the mass of models involved, taken together, be sufficiently adequate and accurate. These myriad models must be mutually:

- correct
- complete
- consistent
- current
- cogent
- coherent

This sort of mass modeling is accomplished successfully hundreds of billions of times per year in modern mass society. Nearly all of the products and services you use every day involve this sort of superintelligent practice modeling.

Managerial superintelligence requires that a vast diversity of models be applied in myriad specialized combinations to support myriad specialized collaborations. Modern mass modeling and mass management systems support and sustain the required mass communication and mass collaboration.

Practical informatics contributes to managerial superintelligence it two ways:

- minimizing infodynamics impedance: lesser managerial modeling effort and expense
- maximizing infodynamics intensity: larger managerial efficiency and effectiveness

## *Practical Prose and Practical Programs*

The practical meaning of practical prose and practical programs is always the exactly the same. Every practical program is an algorithmic restatement that automates and animates some interesting set of practical prose models. The meaning of these prose models is ultimately applied mechanics meaning and applied mathematics meaning.

How can the meaning of practical prose and practical programs be the same? They sure look different!

The meaning is the same because in both cases the practical meaning is constructed from exactly the same expository meaning elements. So the rigor, rationale, and rhetoric elements are the same. Thus the exposition structure is exactly the same. Only the encoding structure is different.

Digging out the empirical exposition structure of practical programs is not always easy. The structure is always there but it's mostly implicit. Techniques of software science reengineering are required make this meaning explicit. This involves much guesswork and groping in the dark.

Even so, this sort of reengineering is what software professionals spend much of their time doing. Every working software engineer spends a good part of every working day recovering and reconstituting the exposition structure of practical software encodings.

Many software professionals will protest that the practical meaning of practical programs is really very different than that of practical prose. In this view the meaning of practical programs has something to do with universal language and universal logic. These logics and languages are aspects of universal computation.

This view is in fact correct, but only for encoding rather than exposition. In any practical application universal computation encodes everything but explains nothing.

Every conventional programming language does in fact support constructions in universal language and universal logic. Yet these languages and logics are strictly encoding languages and logics. When it comes to practical exposition these computational languages and logics are entirely meaningless for all practical purposes.

This meaninglessness is a good thing because it's the genesis of generality. This meaninglessness is the basis for current best practice universal modeling in information technology. General purpose IT universal modeling systems provide a uniform set of commoditized encoding models, methods, and mindsets that work for every sort of subject matter.

This extreme generality is important because of the huge diversity of software application subject matter. This subject matter diversity includes the entirety of the basic, natural, applied, mathematical, and social sciences. Then there are all the fantasy worlds of the entertainment industry just for good measure.

Universal modeling is applicable to all these diverse sorts of subject matter. Without the extreme generality of universal modeling computers might still be laboratory curiosities today.

The good news about universal modeling is that it works for every sort of subject matter. The bad news is that it doesn't work particularly well for any particular sort of subject matter.

So universal modeling is useful for saying almost nothing about almost anything. In practice universal modeling is ideal for high information quantity, low information quality informatics. This sort of modeling is mostly what we do today in practical software projects and packages.

The primary approaches to IT universal modeling in software engineering today are:

- object-oriented[4]: software type structure modeling
- entity-relationship[5]: schemaware type structure modeling
- Universal Modeling Language[6]: architectural modeling

All of three approaches start with trivially universal constructive exposition spaces. From a practical exposition perspective these are metaphysical thingologies.

This is what makes universal modeling universal. Universal modeling begins by throwing away the entirely of applied mechanics and applied mathematics meaning. So all that remains is minimal metaphysical meaning.

In this way universal modeling tames the titanic complexity of practical information by trivializing it rather than tackling it. In end-user informatics the goal is to tackle this complexity by means of understanding the layout and landscape of practical prose language and logic.

All three forms of IT universal modeling are really just very specialized forms of the utilitarian modeling that we all do everyday. In each case the genius of this specialized modeling is systematic subject matter simplification.

Why must we do this systematic simplification? Simplification is necessary because human intellectual gifts and graces are limited. The horsepower in our heads is sufficient to handle only very simplistic subject matter in handcrafted pattern programming.

So the secret to getting software shipped is simplification. The sophistication of real world practical prose models must be systematically stripped away. Simplification is a key success and survival skill in the software world.

In IT universal modeling systematic simplification is accomplished by various means. The randomness of real life is replaced with regularity. Nearly all of the referential richness is stripped away. Redundant perspectives are removed. All of these contribute to loss of content meaning.

Then there is decontextualizaiton. This is loss of context meaning. The native end-user audience context is mostly stripped away. The original context is replaced with a radically simplified artificial context created for the convenience of the software engineering team. This new context usually reflects the very limited subject matter mastery of this team.

By these means systematic simplification strips all of the native "smarts" out of subject matter practice models. So all that is required for smarter software is merely to stop throwing all of these smart parts away. This necessarily requires utilitarian methods of informatics modeling rather than universal methods.

Moreover, systematic simplification strips away most of the end-user informatics value. So practical informatics end-users must settle for the application value they can get rather than the much greater value that they have always wanted.

The mechanical and mathematical meaning that remains after simplification is atomized. This meaning is restated analogically in terms of computational algebras and algorithms. This anecdotal atomization is captured in terms of suitable software and schemaware modeling languages.

Atomization involves mapping practical prose articulation meaning to practical program algorithmic meaning. The mapping preserves the exposition but results in a radically different encoding.

This atomization mapping is constructed entirely as mentalware. In universal modeling there is no alternative by definition. If you can capture the practical meaning of this map as modelware then you are necessarily doing utilitarian software modeling.

Today we don't yet have the practical exposition frameworks or supporting tools to capture this map. If we did then we would be mass custom producing practical informatics software routinely.

One way to understand this atomization map is as a form of encryption mapping. We encrypt information by restating it from familiar exposition and encoding contexts to foreign ones.

Manual methods of software science reengineering are employed to reverse this encryption. These methods start by reconstructing the missing mentalware atomization mapping framework. This reconstruction is always a guesstimate.

Why must we recurrently reconstruct this map? Because this map is mentalware. Mentalware is profoundly lacking in precision, portability, and persistence. It doesn't share well, it doesn't travel well, and it doesn't last very long.

So reconstruction is required even if the original software development team is still available. Often this is not the case. Reconstruction by other teams often turns out to be inordinately difficult. For this reason many reengineering projects turn into rewrite projects.

The atomization map reconstruction in reengineering must always be an exacting estimate. A precise, painstaking reconstruction is required for accurately and adequately decrypting the embedded, encrypted practical prose meaning hidden in every practical program.

If the meaning of practical prose and practical programs is identical then why is it so hard to produce software? Why do remarkably smart people have to do lots of really hard work in order to get even the simplest practical informatics software shipped?

The answer is wasted mental motion. All manual modeling involves substantial mental motion wastage. The tactical benefits of all forms of model automation always involve radical reductions in wasted mental motion.

It turns out that universal modeling methods are uniquely wasteful. Universal modeling involves many orders of magnitude more wasted mental motion than equivalent practical prose utilitarian modeling.

This is hardly surprising. The generality of universal modeling comes at a price. That price is the extreme mental effort required to get from general to genuine in any particular practical application.

Universal modeling in information technology is intellectually intensive in the extreme. At the same time this sort of modeling is exceptionally intellectually inefficient and ineffective.

Intellectualizing a single universal modeling module typically involves at least five very different kinds of markup and mentalware models. Much routine software engineering involves simultaneously intellectualizing several different modules at once.

This is not some extreme feat of software engineering. This is how practical software developers spend much of their working lives. So it's a wonder that their heads don't explode more often.

In addition to extreme complexity there are two other major wasted mental motion sources. These are mentalware modeling and metaphysical pattern modeling.

Any sort of modeling that involves complex mentalware models is wasteful of mental motion in the extreme. Complex mentalware is always involved in pattern programming projects.

Moreover, pattern modeling of any sort is always far more wasteful of mental motion than equivalent principled modeling. This is just one of many reasons why other practical pursuits abandoned pattern modeling decades ago.

Given all these inefficiencies it's no wonder that automating and animating fairly simple practice models can become a Herculean effort. All of these inefficiencies are built into IT universal modeling. They can't be fixed because they aren't broken. They are built in at the foundations.

Another major problem unique to universal modeling is model reuse. All mature modeling practice starts with prior models. You next model is very much like your last one. There are extensive stocks of precedent and prototype models that make it unnecessary to reinvent the wheel in each new modeling application.

This is seldom the case in universal modeling. In most cases every new universal modeling effort invents some slice of practical reality from scratch. All of the application applied mechanics and applied mathematics meaning is carefully constructed from the ground up.

The result is a completely isolated *de novo*, disposable Informatics worldview. This is a closed, self-contained, self-sufficient application context.

All this doesn't seem to make much sense. All the applied mechanics and applied mathematics models must have some prior application somewhere. So why reinvent new ones rather than just reusing old ones?

The answer is that reusing existing software models requires reengineering these models. You can't risk reusing these models unless you thoroughly understand them. You can't thoroughly understand them unless you thoroughly reengineer them. This reengineering is almost always far more expensive than just rewriting the same models again from scratch.

Note that all of this IT universal modeling is utterly redundant. It is merely a restatement of practical exposition from the communal utilitarian modeling context to a unique new universal modeling context.

Is all this really necessary? Today it is, but in the age of end-user informatics it will be unnecessary.

We will do something much more reasonable. We will just synthesize practical programs directly from practical prose. Mass custom end-user informatics software will become an automatic byproduct of the routine managerial modeling we must always do anyway.

All this is not in any sense revolutionary or even new. All of the various technologies required are commercially mature with one exception. That one exception is mature models of the technostructure of practical prose. This is the one big barrier to commoditizing, commercializing, and consumerizing mass custom end-user software production.

Hopefully we will come by these better models sooner rather than later. It is these models that will make end-user Internet informatics a commonplace reality.

This is do-it-yourself, demand-driven, disposable informatics. This is just the informatics solutions you need just when you need them.

## *Practical Information Quality and Safety*

Practical information quality is about:

- modeling exactly everything you mean
- meaning exactly everything you model

The elements of practical information quality are:

- interoperability: meaning that works well in model mashups
- insurability: meaning that reliably means what it means
- immortality: meaning that does not disappear over time

Practical information quality and safety are closely related issues. Information safety is always a question and information quality is always an answer.

Information safety is about managing informatics communication and collaboration risks. It's about making sure that an order for a hundred cases of brake shoes doesn't put a hundred cases of ballet shoes on the loading dock.

There are lots of different kinds of information safety risks. These are entirely practice dependent. So there are no fully general ways of managing information safety, just lots of genuine ones.

Information quality cannot be managed unless it can first be modeled. So information quality requires explicit models and metrics of practical meaning quality.

Manageable information safety and quality cannot be achieved with any form of universal modeling. The implicit meaning of universal modeling mentalware and metaphysics is intrinsically unmanageable. This is why there is no workable way to model information quality and safety in mainstream informatics practice today.

Manageable information safety requires utilitarian modeling mechanistic meaning contexts. Any feasible approach to automated information safety management will require these contexts.

The starting point for information safety is interoperability. Without interoperability the issues of insurability and immortality are moot.

The degrees of informatics information safety are:

- plug-and-play: free, first try fit, fail-safe, fully automatic impromptu interoperability
- plug-and-pray: hack it all together and hope for the best
- plug-and-pay: do lots of recurrent reengineering

In manufacturing engineering an important distinction is:

- first-try-fit: parts that should fit together always do
- file-to-fit: some part modifications are usually required

This same sort of distinction applies in informatics. An important part of plug-and-play informatics is first try fit informatics components. This is always necessary but never sufficient. Free combinability, fail-safety, and full automation are also required.

Today the interoperability of independent informatics installations is almost always plug-and-pay. In these installations plug-and-play is possible only for the most trivially simple sorts of informatics models.

In nearly all cases maintaining interoperability between any two independent informatics installations is a remarkably costly and chancy undertaking. This seldom succeeds for more than relatively simple sorts of information flows.

Achieving reliable interoperability among independent informatics installations requires lots of recurrent reengineering. This may also require manual support systems of various kinds.

Manual software and schemaware model conciliation is the starting point for recurrent reengineering. In addition there are usually some manual backup procedures for fault recognition and recovery. There may also be a need for manual workaround procedures for cases that the informatics can't handle. These are all continuing everyday expenses.

Then there is informatics catalog and content reengineering. This is another kind of very costly recurrent reengineering.

This sort of reengineering is about the conciliation of database models across independent informatics systems. This sort of reengineering involves making sure that part numbers, prices, packaging codes, and lots of other minute details are meticulously matched up.

This usually involves maintaining manual mappings between multiple sets of code lists and naming conventions. This is often done on a weekly or monthly basis. In some cases changes must be made on a daily basis.

How expensive is all of this recurrent integration and interoperation work? Consider the case of full end-to-end EDI (Electronic Data Interchange) integration of two procurement chain informatics systems. Specifically assume a single fully automated buy-side to sell-side EDI linkup.

This is the most expensive and elaborate way to do EDI. The exchange of a dozen or more types of EDI massages is required.

A conservative rule of thumb in large manufacturing installations is that maintaining this sort of link costs ten thousand dollars per year per message per link. So hundreds of thousands of dollars per year is required to maintain limited EDI interoperation of just one pair of independent informatics systems.

Skimping on these expenses can lead to major incidents of information sorrow. This is how brake shoes turn into ballet shoes.

Why is all this effort and expense required to make informatics systems work reliably in tandem today? The reason is the sort of universal modeling done in best current practice IT.

In universal modeling the mechanistic meaning of practical information is implicit rather than explicit. So recovering this practical meaning necessarily requires recurrent reconstruction of explicit meaning by means of laborious reengineering methods. There is no room for error here.

This isn't exactly computer science work. It's more like computer séance work. It's about resurrecting and recovering missing meaning. It's about trying to divine and deconstruct the implicit mentalware mechanistic meaning context of complex software and schemaware models.

This is why informatics reengineering, done well, is remarkably expensive. Informatics reengineering done poorly is far more expensive.

Small informatics reengineering errors can be ruinously expensive. Major reengineering errors can put you out of business.

Informatics catastrophes are far more frequent that most companies would like to admit. In nearly all cases these are due to major reengineering errors.

Because of universal modeling we must spend huge sums of money on recurrent informatics reengineering. Something on the order of one hundred billion dollars worth of recurrent IT reengineering projects are done worldwide annually by commercial IT service companies. The cost of captive reengineering is unknown but probably amounts to several times this again.

These vast sums are not in any sense accounting fictions. They are real redeemable cash flows. They are just one way of understanding how expensive universal modeling actually is in practice. Note that no other practical modeling discipline engages in anything like this sort of recurrent reengineering.

Someday manageable information safety and quality will eliminate all this unnecessary expense. This will first require replacing universal modeling informatics practice with utilitarian modeling practice. This is how we can, must, and eventually will achieve routine plug-and-play interoperability in practical informatics.

Some elements of practical informatics plug-and-play include:

- polygamy: everything works with everything
- profligacy: everything works together all the time
- polyglotism: everything speaks all languages
- promiscuity: lots of choices for working together
- profluency: everything works effortlessly

## *Practical Meaning*

Meaning explains observable order. The history of practical ideas is about mankind's search for the meaning of natural order. It is about our efforts to create artificial order out of natural order.

Meaning is mandatory. People can't cope with meaninglessness. So every sort of observable order must be made meaningful.

We always use the best models we have to understand the meaning of observable order. At the dawn of humanity our best models were mostly myth and magic models. In modern mass civilization our best models are applied mechanics and applied mathematics models.

In human prehistory we understood thunderstorms as the work of the weather gods. Today we understand thunderstorms in terms of thermodynamics, fluid dynamics, and electrodynamics.

There are three alternative ways of understanding the meaning of observable order. These include mechanical, mathematical, and metaphysical meaning. Each of these three sorts of meaning is entirely different and distinct. Sometimes they are complementary.

The three basic sorts of meaningful order are summarized as:

| Meaningful Order | Champion in Greek Antiquity | Truth | Source |
|---|---|---|---|
| materialist | Protagoras | empirical | rationalization |
| mathematical | Pythagoras | exact | reflection |
| metaphysical | Plato | esoteric | revelation |

Metaphysics is an indispensable part of any balanced personal worldview. Will time end tomorrow after breakfast? There is no hard science evidence to the contrary. Even so, we keep on making plans for lunch.

Nevertheless, metaphysical meaning plays no significant part in practice modeling today. Practice modeling always combines an appropriate balance of mechanical and mathematical meaning.

Plato has always had better publicists than Protagoras. Yet practical people everywhere are disciples of Protagoras rather than Plato today. They have statefull, scientific, sophisticated working worldviews. They have no time or tolerance for universals, metaphysical ideals, exact truth, or absolute truth in any form. They are adherents of utility, mechanistic meaning invariants, empirical truth, and approximate truth.

Mechanical order is the actual order of physical reality modeled ultimately in terms of the shapes that characterize physical state. Mathematical order is abstract order derived ultimately from constructions on the natural numbers. Metaphysical order is everything else.

Perhaps there is a god of sunsets. If so, we will never demonstrate the metaphysical existence of this god with telescopes, microscopes, and such. Metaphysical order is always necessarily occult rather than observable. All of the observable order in physical reality and practical reality is materialist and mechanistic.

Notions of veracity and validity, of truth and proof, are very different in mechanical and mathematical modeling. In mechanical modeling truth is approximate and proof is accomplished by means of the scientific method. In mathematical modeling truth is absolute and accomplished by means of mathematical proof theory.

Practical modeling always employs an appropriate balance of mechanistic and mathematical meaning. In practical modeling mathematics is always a tool rather than a truth. The truth of mechanistic systems lies entirely in physical state dynamics. Applicable mathematical patterns provide useful approximate analogs of actual and abstract physical state and shape dynamics.

Approximation is unavoidable in practical modeling. The totality of the physical information contained in just a single grain of sand is many orders of magnitude beyond the capacity of all the world's computers combined.

Practical modeling is always parsimonious in the sense that approximations are always chosen to suit some specific range of intended applications. This is the law of least articulation.

It is said of some people that they will tell you how to build a clock if you ask them the time of day. This serious social *faux pas* is an egregious violation of the law of least articulation.

In practical modeling much more information is always discarded than is detailed. Success in practical modeling always starts with a carefully considered choice of approximations. Muddling is excessive modeling. In practical pursuits muddling is always a sign of modeling immaturity.

## *Practical Measurement*

In practical pursuits measurement limits meaning. If you can't measure a practice you can't, for all practical purposes, mandate the practice. Thus you can't manage the practice.

Note that this disposes of most of the thorny problems of classical epistemology. But does this substantially or even severely limit the range of practical meaning that can be modeled?

Today we're pretty good at measuring most kinds of physical and practical meaning. Even so, there are times in some practical pursuits when policy trumps physics and practice.

Consider the notion of reasonableness which is critical to systems of jurisprudence based on the English Common Law. In this system juries decide the facts while jurists decide the law.

In criminal cases juries are expected to make determinations of fact beyond all reasonable doubt. This is the "reasonable man" standard.

Now consider a murder trial where the verdict hangs on whether the defendant intended to murder the decedent. The jury is expected to hear the evidence and decide the issue.

Of course none of the jurors could possibly know the mental state of the defendant at the time that the murder was committed. So they must make what amounts to a subjective judgment on this issue. Yet criminal cases of this sort could not be adjudicated without some means of sorting out facts of this sort. There is no substantially more objective way to decide the issue.

Once the jury made it's decision the issue, apart from exceptional circumstances, is settled. This issue cannot be reviewed on appeal. Appellate courts may, if fact-finding was defective, refer the case back to the trial court for consideration by another jury.

Failing a finding of trial court error the facts decided by the jury are true as a matter of policy. This means that they are subsequently true for all practical purposes.

This sort of policy truth is sometimes used in lieu of physical truth or practical truth. Many practical pursuits use models with meaning that is unmeasured on not measurable. Commonplace examples include accounting estimates, legal fictions, and educated guesses. These are used either because they are simply convenient or because there is no other practical alternative.

## *Practical Truth and Proof*

What is truth? For all practical purposes truth is whatever works. There is no absolute or exact truth in practical pursuits. The working word gets along just fine with approximate, empirical truth estimations.

The two primary sorts of truth are:

| Perfect Truth | Practical Truth |
|---|---|
| exact | estimation |
| universal | utilitarian |
| absolute | approximate |
| paramount | pluralistic |
| stateless | statefull |

If you need perfect truth then go see a mathematician or a metaphysician. But don't expect the truths provide by these folks to be of much practical use *per se*.

Physical truth is defined by state mechanical models and proved by means of the scientific method. Physical truths are those that are verifiable, falsifiable, and verified by physical means.

Perfect truths models, on the other hand, are stateless. Perfect truth proofs are accomplished by means of mathematical or metaphysical proof theories.

What is a chair? This may seem to be a simple enough question. Even so, perfect truth and practical truth provide very different answers.

The perfect truth chair is a Platonic form composed of a structure of universal ideals. Legs, arms, and such are examples of these ideals. There might be many different kinds of chairs, but they all occupy the same category of forms. They can be demonstrated to belong to this ideal category because they share a common construction of ideals.

The practical truth chair is something you can reasonably sit on. So for practical proof purposes almost anything might serve as a chair. Some candidates include a rock, another person, or the throne at Buckingham Palace. Even a block of ice might do if you don't plan to sit for too long.

Proving the practical truth of a chair is accomplished by sitting. This is a very simple form of the scientific method. If sitting works then you have the best proof there can ever be of the practical truth of your chair.

## *Practical Models*

How can we model the working world? For physical and practical modeling purposes there are only two basic choices:

- metaphysical patterns
    - theology: abstract metaphysical patterns
    - tradition: applicable metaphysical patterns
- mechanistic principles
    - technology: applicable mechanistic principles
    - theory: abstract mechanistic principles

We are all pattern prodigies. We see patterns everywhere in everything. We strive to make sense of these patterns in whatever ways we can.

Much of modern science evolved from pattern pseudoscience and protoscience. Astronomy evolved from astrology. Number theory evolved from numerology.

Today we understand that we live in a universe prone to spurious patterns. There are whole branches of materialist and mathematical science that explain the provenance of all these spurious patterns.

For most of human history the best practice models available have been pattern models. Consider the case of Hippocratic medicine.

The Greek physician Hippocrates (460-370 BC) pioneered a system of medical practice based on physiological patterns termed humors. This system, known as humorism, was widely employed in medical practice up until the end of the 19th century.

In humorism every human health issue can be understood in terms of four fundamental humors. In this system every health issue can be modeled as some specific balance of the blood, black bile, yellow bile, and phlegm humors.

Today no reputable physician would dream of using humorism in medical practice. This sort of pattern practice model, the state of the art for many centuries, has been entirely supplanted by modern physiological principles.

In just this way principles have replaced patterns in nearly all practical pursuits over the course of the last hundred years. In this way the fundamental nature of practical knowledge has changed drastically.

One way to think about this change is the following distinction:

- metaphysical patterns
    - know-how practical knowledge
    - metaphysical craftwork modeling
    - doing what you know
    - judgment without justification
- mechanistic principles
    - know-why practical knowledge
    - mechanistic clockwork modeling
    - knowing what you do
    - judgment with justification

In the modern industrial world principles are the standard for nearly all sorts of practice modeling. Principles are always the legal standard for principled professional practice.

Professionals must always be prepared to justify their judgments in terms of best practice mechanical principles. This is the legal standard today throughout the industrialized world.

Patterns, of course, have not gone away. Today the role of patterns has changed radically in practical pursuits. Patterns once were answers but now they are questions. Patterns are questions which we always answer by the application of proven principles.

Consider a visit to your doctor for some malady. Your doctor will model the condition of your health in terms of medical signs and symptoms. These are physiological patterns which must be matched to the correct principled pathology model. Various sorts of tests and specialized examinations may be required to sort out the right model.

Abstract patterns have always been the basis of mathematics. G. H. Hardy famously teaches us that mathematics is the science of patterns.

The thing that we must always remember is that patterns alone aren't practical. Applied mathematics always uses mathematical patterns as a way to encode mechanical principles. These combinations are always analogical. They are analogies between the shape of physical state and a geometric interpretation of some suitable mathematical structure. Mechanistic models are always approximations so the result is an analogy to an approximation.

So good practical modeling always requires a suitable balance of mathematical patterns and mechanistic principles. Today many powerful mathematical pattern models go begging for lack of suitable matching mechanistic principled models.

One example of this is game theory. Game theory is a very powerful family of formalist models in decision theory. Game theory has lots of potentially advantageous applications. Game theory ought to be far more widely applied than it is in management support software applications like business intelligence.

The reason that game theory is not more widely applied is information quality and information safety limitations. Workable game theory modeling requires high information quality and safety game models. Typical sorts of IT models aren't nearly good enough. So high quality game models must be laboriously assembled from diverse documents and databases. Game models must be checked and rechecked for information safety.

This involves a lot of effort and expense. Moreover, it builds in a lot of decision lag time. Thus game theory today finds only sporadic practical application in management decision making.

What if mechanics is really mathematics in disguise? This notion is the basis for ultraformalism. Ultraformalism aims to reduce the entirely of universal order meaning to mathematics.

Some folks should be ultraformalists. If you're a wizard of mathematical physics like Roger Penrose then ultraformalism is the *Road to Reality*[7]. Even so, ultraformalism is the road to ruin for practical people in nearly all practical pursuits. In these pursuits an appropriate balance of mechanical and mathematical modeling must always be maintained.

Ultraformalism is the basis for the notion that universal modeling is ultimately perfectible. In this view perfection will be achieved by discovering sufficiently powerful formalisms. These formalisms must be mathematical patterns that reduce all of mechanics to some convenient, canonical-closed, constructive modeling space.

In this way patterns can stage a comeback. Pattern meaning can be pushed to parity with principled meaning and then onwards to perfection. Approximate empirical meaning will be obsolete. Absolute exact meaning will become obligatory in all practical pursuits.

In this way practical information will be reinvented. Modern mass civilization will be thoroughly remodeled in the most literal sense.

We may even discover a universal, unified, ultraformalist perfect pansophical science of everything. We might just discover the language and logic of ultimate universal meaning. Today, after three thousand years of searching, we are still in absolutely no danger of making these discoveries.

Perhaps someday we actually will find magic formalist patterns. There has never been a shortage of candidates. Theoretical computer science has always been a prodigious source of promising new formalist patterns.

This is why we have lots of remarkably useful computational pattern models today. These find diverse applications in a great variety of practical pursuits. Business intelligence, statistical demographics, Internet search engines, machine transliteration, and image recognition are just a few examples.

Even so, patterns always complement principles in practical pursuits today. Patterns are not nearly competitive as a primary means of practice modeling. Nor is it likely that patterns will become competitive with principles in the foreseeable future.

Surprises, of course, are always possible. New formalisms and new applications of formalisms appear constantly. Mathematical unification programs, such as the Langlands Program[8], are uncovering lots of previously unsuspected relationships among families of mathematical patterns.

On the other hand Gregory Chaitin argues persuasively[9] for substantial intrinsic limits to the power of mathematical reasoning. It just might be that the phenomena structure of pure mathematics is highly randomized just as the phenomena structure of pure mechanics.

Note that most pure mathematics deals with infinite structures. The natural numbers (all positive integers) is such a structure. The usual methods of mathematical proof theory[10] work well on these infinite structures.

The models of practical informatics are typically finite for the purposes of mathematical reasoning rather than infinite. For many purposes finite mathematical model theory is substantially less powerful than infinite model theory. An important but rather obscure point is the failure, in general, of the compactness assumption. A classic paper[11] by Ron Fagin provides a mathematical introduction to limitations of finite model theory.

Finite model theory limitations may restrict the range of mathematical reasoning applicable to typical practical informatics models. For example the appropriate analog of the classical mathematical proof strategy of induction may turn out to be brute force search.

Today there is no verifiable evidence whatsoever for any intrinsic relationship between mechanistic meaning and mathematical meaning. There are endless conjectural relationships but so far these are all counterfactual. None of these conjectures has ever been definitively demonstrated. Moreover, there seems to be no danger of any such demonstration in the foreseeable future.

Any unification of mechanical and mathematical meaning must preserve the correctness of the mechanistic foundations of the hard sciences and high technologies. Otherwise airplanes will fall from the sky.

Moreover, any such demonstration must show that these foundations are substantially incomplete. In addition any useful demonstration must also show some very substantial simplification of our materialist order models. Of course any demonstration must also satisfy the standards of the scientific method.

Until this demonstration is an accomplished fact we will do well to assume that metaphysical, mathematical, and material meaning are three uniquely different kinds of meaning. In some cases these may be complementary but they should never, ever be confused.

Ultraformalism systematically confuses mathematical and material meaning. This confusion is a rich source of formalist innovations. This innovation serves to provide us with lots of new and potentially useful modeling patterns.

Even so, these patterns cannot provide a path to strategic informatics progress. In decades past practical informatics progress was primarily limited by our understanding of pattern encoding models. Today practical informatics progress is primarily limited by our understanding of principled exposition models. This will likely remain true for many decades to come. This may well remain true for centuries to come.

Thus principles, not patterns, are the path to strategic informatics progress today. Strategic informatics progress requires that we understand the mechanistic principles of practical pursuits. Strategic informatics progress requires that we understand the principled basis of utilitarian modeling. Strategic informatics progress requires that we understand how practical people use principled information in practical pursuits.

# 2: The General Theory of Practical Collaboration

## *Introduction*

You are collaborating with this author right this very instant. This collaboration is possible because we share a working worldview in common. That's why these working words make sense.

Who makes and maintains all these working worldviews? How do we use these working worldviews? How can we best automate and animate these working worldviews? These are the questions answered by the General Theory Of Practical Collaboration.

## *Working Words and Working World Models*

Somehow we all routinely manage to model the working world with working words. We use many thousands of these working words every day to communicate and collaborate in our various practical pursuits.

How do we do this? Sometimes this sort of seems like magic. But as with all workable magic the secret is high mechanics.

Working words are meaningless in and of themselves. The significance of working words always starts with the context of a working worldview in some working world practice community.

The working world is the world of practical pursuits. There is, of course, a great diversity of practical pursuits. Every sort of specialized practice is supported by a practice community. These practice communities maintain the shared communal context of practice communication and practice collaboration.

Today there is a huge diversity of specialized practice communities in the working world. Each of these communities supports some range of practice mechanisms. We must explore the working world in order to understand what these mechanisms are and how they are modeled.

Each specialized practice community has its own working worldview. Your home has one, your place of employment has one, and your neighborhood has one. These worldviews are all different but there is much commonality among them.

A working worldview is a set of working world models. Each working world model captures the mechanistic meaning of some working world practice mechanism. Each working model supports some range of significant working word constructions.

These working models are expressed technologically in terms of practical prose. In this sense practical prose can be understood in terms of tools, toolkits, and techniques.

In this sense every working word is a little tool. Each working word maps to some particular sort of specialized working world model meaning.

Tools come in toolkits. Working world models are specialized toolkits. Each of these shared reference models supports flexible working word constructions.

Each construction captures the mechanistic meaning of some specific working world practice mechanism. These constructions are the techniques of practical prose technology.

## *The Working World and Working Worldviews*

How many people does it take to change a light bulb? These days it takes thousands of people. Every time you change a light bulb you enlist the aid of a great diversity of practical people. These people are engineers and economists. They are line workers and logistics analysts. They are mining and manufacturing experts. Then there are many, many more besides.

All of these different folks communicate and collaborate in different ways during the course of helping you change your light bulb. You will never meet most of these people. Even so, you couldn't possibly change your light bulb without their help.

How do all these people manage to communicate and collaborate so well? They share common working worldviews. These always include the everyday working worldview as well as various specialized elite working worldviews.

Practical people everywhere share a common consensus everyday working worldview. This working worldview is a collection of everyday working mindsets, models, and methods.

This everyday working worldview is highly pluralistic. It supports many alternative and complementary practical reality perspectives. A few of the most important of these practical perspectives include:

- artifact and asset
- control and commonality
- duty, domination, and discipline
- mechanism and instrumentality
- resource and reciprocity
- management and enterprise

This pluralism of overlapping practical perspectives is highly redundant but hardly superfluous. It is exactly this reinforcing pluralism that provides the power and precision of the common consensus working worldview.

Every working worldview, including yours, is a set of shared mechanistic models. These models are about how we survive and succeed in the working world. These models are about:

- working wisdom: models of how the world works
- working wile: models of how to work in the world

Where does the working worldview come from? We all contribute to the common everyday working worldview consensus. It is the massive collective competence of practical people everywhere that serves to order and optimize the common consensus everyday working worldview.

We all use these practical models every day. Everyday examples include road maps, cookbooks, broadcast schedules, financial market prices, and so on. Examples of elite models include specifications of automobile designs, statistical analyses of pharmaceutical product trials, and macroeconomic trends in Central America.

Where do our working models of practical reality come from? We've been accumulating these models since the dawn of agricultural civilization about ten thousand years ago.

Most of the models that we use today are more recent, generally less than one hundred years old. Even so, some are much older. Double entry bookkeeping dates back to $11^{th}$ century Venice. Many of the most useful working words have their origins in Greek and Roman antiquity.

## *The Elite Working Worldviews*

Elite practical prose modeling is big business today. About three hundred million talented folks worldwide do about three trillion dollars worth of elite practice modeling. These folks are managers, professionals, technicians, and specialists. Most have extensive specialized education and experience.

The infrastructure in place to support elite practical modeling has a replacement value on the order of one hundred trillion dollars. This includes modeling stocks, staffs, standards, and systems.

How many elite working worldview are there? That depends on how you count. Everybody in the working world has at least one, and many of us have several.

The basic everyday working worldview seems largely to be the same throughout the industrial world today. There are, of course, international, industrial, institutional, and individual differences. These are apparently minor.

This means the practical exposition structure of all of the world's major industrial languages is essentially the same today. If this were not the case then some practical pursuits would be intrinsically advantageous in areas where some of these languages are spoken.

This is not apparently the case. The reason for this is that the exposition structure of practical prose is extremely plastic. It constantly evolves to accommodate applied mechanics and applied mechanics advances.

Even so, practical English does seem to have something of an edge. Practical English has become the worldwide *linga Franca* of many practical pursuits. Examples include aviation, telecommunications, some medical specialties, and several engineering specialties.

Practical English may very well have an unusually rich and redundant exposition structure. This is would not be surprising given its polyglot origins.

Moreover, practical English is the mother tongue of both the industrial revolution and industrial modernism. So it's not surprising that other languages frequently borrow exposition elements from practical English.

## *Working Worldview Context Communities*

Elite working worldviews are shared by institutional modeling communities. Each modeling community has its own specialized context and concerns. You neighborhood bakery has its own communal context.

Elite communal contexts are largely constructed from combinations of specialized topical and target contexts. Each of these specialized contexts is shared in common across participating communities. Those that use these specialized contexts form topical and target context communities. It is these communities that maintain the context over time.

How many practice context communities are there? How many topical practice contexts are there? How many target practice contexts are there?

It would be convenient if there were a complete master plan and the complete master parts catalog for modern industrial civilization. Regrettably, these don't exist.

But what if they did exist? What would they tell us about the complexity of practical information contexts? What would we find by measuring the volume, variety, and value of all these models?

In fact nobody really knows. This is just one of many vitally important things that we don't know about practical information.

The only thing we do know with any certainly today is that the model stocks supporting modern industrial civilization are huge and heterogeneous.

So let's make an educated guess. For present purposes assume that there are hundreds of thousands of topically specialized practice modeling varieties. Further assume that there are hundreds of millions of target-specialized practice modeling varieties.

One example of a topical context community is industrial sugar. Industrial sugar comes in a great variety of forms. It is produced, packaged, and put to use in a wide variety of ways. There are many international sugar standards as well as national sugar laws and sugar regulations in most industrialized countries.

Those that make and use industrial sugar need a common context to sort all this out. This includes sugar refiners, sugar brokers, packaged food companies, food service companies, food retailers, regulatory agencies, and so on.

So the world of industrial sugar has industry organizations, trade organizations, technical organizations, standards organizations, research institutions, industry publications, industry conventions, and so on. All of these contribute to maintaining the topical context of industrial sugar logic and language.

## *Working Worldview Evolution*

The working world and the working worldview evolve jointly over time. Mostly this evolution is smooth and steady. Every now and again this evolution makes big leaps.

Our common everyday working worldview underwent a titanic transition in the first half of the $20^{th}$ century. This was the transition from the industrial revolution working worldview to the industrial modernism working worldview. This was the historic transition from a craftwork to a clockwork consensus working worldview.

This transition created the working world as we know it today. For present purposes this transition will be termed the Great Remodeling. This is clearly one of the all-time great technology transitions in human history. It's right up there with fire and the wheel.

Remarkably, this historic transition is mostly unknown to historians of the $20^{th}$ century. These experts well understand that the practical culture of the industrial world changed radically. Yet the pivotal role of practical language and practical logic in this transition seems to have escaped their notice.

Thomas Edison[12] is the prophet of the Great Remodeling. Like all great $19^{th}$ century inventors Edison was a brilliant tinkerer. Edison's genius lay in his talent for systematic tinkering. He popularized the notion that a systematic understanding of the mechanics of practice could provide the means for deliberate, deterministic, disciplined innovation and invention.

The Wright Brothers and Henry Ford are among the most notable of Edison's many disciples. Ford was actually an Edison protégé, serving for a time as Chief Mechanic at the Edison Electric Generating Company. Ford and Edison remained fast friends throughout their lives.

Industrial modernism is the ideology of the Great Remodeling. Henry Ford pioneered and tirelessly promoted this ideology. It was Ford who famously proclaimed the machine as the new messiah[13].

In many ways Ford's industrial modernism is Edison's emphasis on systematic mechanics evolved into an industrial ideology. Industrial modernism holds that everything can, must, and will be advantageously mass produced. In this view the key to advantageous mass production is a thorough understanding of mass practice mechanics.

Ford was right. Today nearly everything in the modern industrial world is advantageously mass produced. This is radically different than the way that things were produced just a century ago.

Moreover, mass production continues to evolve. New kinds of mass production emerge. Mass commodity production evolves toward mass custom production.

Today industrial modernism is the basis for the working worldview throughout the modern industrial world. This relatively recent clockwork working worldview is radically different from the craftwork worldview that prevailed just a century ago.

The craftwork working worldview was pervasive in all human civilizations from the dawn of agricultural civilization until the early twentieth century. The craftwork working worldview is a metaphysical worldview based on theological and traditional patterns.

The modern clockwork working worldview is based on mechanistic technological and theoretical principles. This worldview is mandated and maintained in all principled practical pursuits today. It is required both by common sense and by corpus juris.

Today this is the working worldview of managers, professionals, technicians, and specialists in all principled practical pursuits. It is particularly the worldview of the principled professions. Professions of this sort include law, accounting, engineering, economics, finance, and so on.

The transition from craftwork to clockwork working worldviews mainly took place between the end of World War I and World War II. It was this transition that enabled and engendered modern mass systems and modern mass society.

Early mass system pioneers understood that they could succeed only by replacing craftwork practice with clockwork practice. Initially the challenge was doing this for mundane making and moving systems.

This first phase of the Great Remodeling gave rise to mass manufacturing plants, mass distribution systems, mass communications media, and so on. Early forms of these systems were pioneered in the 1920s and perfected in the 1930s.

It soon became clear that mass mundane systems demanded radically new kinds of mass management systems. This gave rise to the second phase of the Great Remodeling.

This second phase started towards the end of the 1920s and continued throughout the early 1950s. So most mass management mechanisms in use today date from the second quarter of the 20$^{th}$ century.

The transition from the craftwork working worldview to the clockwork working worldview was highly contentious. This was a major social transition as well as a technological one. The craftwork working worldview did not go quietly.

The industrial science movement was founded to help this transition along. Industrial science started at the Ford Motor Company and rapidly spread to most of the pioneering mass systems industries.

Industrial scientists were drawn from the ranks of factory workers and foremen as well as the ranks of salaried employees and managers. These folks served as the shock troops of the industrial modernism movement. They worked frantically to abolish craftwork in all its forms. These were the folks that Henry Ford meant when he famously said "I am looking for a lot of men who have an infinite capacity to not know what can't be done."

The 1920s saw some early successes in mass systems. There were successes in automotive and appliance mass manufacturing. There were successes in magazine publishing and radio broadcasting. These early successes financed much of the roar of the Roaring Twenties.

Even so, there were far more failures than successes. Thus many initially saw mass systems as nothing more than a fad. The stock market crash of 1929 can be seen as the bursting of a mass systems speculative bubble.

The watershed moment in the emergence of the clockwork working worldview was the opening of Ford's River Rouge plant at Dearborn in 1928. This massive complex provided the definitive vindication of Ford's industrial modernism ideology. People from all walks of life the world over came to River Rouge and marveled.

River Rouge demonstrated complete lifecycle production the automobile. Cars were the most complex mass manufactured product of the day. Sand, rubber, iron ore, limestone, fuel, water, cloth, and wood went into the plant. Cars came out. Clockwork methods rather than craftwork methods were employed at every step.

Today vast integrated industrial complexes are commonplace. In 1928 a visit to The Rouge was a life-changing experience. Every visitor bore witness to the birth of a radically new way of working in the world. Some saw this new way as fortuitous, others were fretful.

Henry Ford initially saw industrial modernism as the key to permanent peace and plenty. He would live to be sadly disappointed in this hope as World War II broke out. This was the first mass systems war in world history.

During the war all of the wartime production authorities in all of the warring countries demanded clockwork production. Vast clockwork production evangelism, education, and enforcement systems were established by these wartime production authorities in every warring country.

Thus by the end of the war the clockwork working worldview was firmly entrenched throughout the industrial world. The craftwork worldview played very little part in the rebuilding and retooling that followed the war. Today the craftwork working worldview is all but extinct throughout the industrial world.

The working worldview that is emerging today worldwide is ultramodernism. The ultramodernism worldview is the working worldview of global cutthroat capitalism.

Ultramodernism is the view that we can understand the working world entirely in terms of elementary and emergent mechanics. Ultramodernism can be seen as an evolutionary extreme of the industrial modernism worldview.

Ultramodernism is the final rejection of the western metaphysical tradition in the working world. This is hardly remarkable since much of the working world today is found in regions where this tradition is far more foreign rather than familiar.

Some, especially in academia, will warn that this rejection is dangerously naive and naively dangerous. In this view substantial mastery of metaphysics is indispensable to any mature understanding of ourselves and of others. This may indeed be true for many purposes but it apparently isn't true for purely practical ones.

So, in the working world at least, Protagoras has triumphed over Plato. The working world is sophisticated rather than Socratic. Mankind has indeed become the measure of all things in practical pursuits.

## *Working World Context Frameworks*

The starting point for deconstructing practical prose meaning is consensus context. Every sort of practical prose modeling is done in the context of one or more practice modeling communities.

The context of a modeling community involves one or more community consensus modeling spaces. It is the lexical and logical conventions of these context spaces that support the interchange of practical model meaning among community authors and audiences.

A context framework serves as a sort of a map of a modeling space meaning spectrum. A commonplace restaurant menu is an example of simple specialized context framework. A menu supports the combinatorial construction of a range of restaurant meals from entrees, deserts, and so on.

Primary varieties of practical context frameworks include:

- cookbook: specific modeling practice prototype collections
- casebook: specific modeling practice precedent collections
- constructive: collections of design principles that are combined into design practice

The best documented examples of institutional intelligence contexts are consensus principled professional practice standards. Some examples typically found in U.S. community libraries include:

- enterprise: Uniform Commercial Code (UCC)
- economic: Generally Acceptable Accounting Principles (GAAP)
- engineering: International Building Code (IBC)

Each of these framework examples are combinations of principles that allow the constructive design of a great diversity of practice models. Auditors, building inspectors, and jurists act as critics of models constructed in the context of these frameworks.

Any context framework can be characterized in terms of coverage. Varieties of coverage include:

- canonical
    - canonical-closed: all reasonable practice cases are covered
    - canonical-close: nearly all reasonable practice cases are covered
- commonplace: most or all routine practice cases is covered
- cursory: only some small fraction of reasonable practice cases are covered

Practice cases may be characterized as:

- commonplace case: everyday practice
- corner case: exceptional practice
- challenge case: extreme practice

The range of design space coverage is illustrated in the preceding examples of constructive frameworks. GAAP, at least theoretically, is canonical-closed. It provides an authoritative basis for constructing the accounting policies and practice of any specialized enterprise. Note that there are various national and economic segment specializations of GAAP.

The IBC is canonical-close. Individual jurisdictions specialize IBC for local circumstances. Even so, it's expected that corner cases will arise. Corner cases are reasonable design cases that would otherwise not conform to the IBC. Planning authorities issue variances when reasonable corner cases arise.

The UCC is cursory. It mostly serves as a fallback for contract contingencies not considered by the parties to a commercial agreement. It is largely a catalog of remedies for routine varieties of commercial fraud, foolishness, and failure.

## *Principled Professional Working Worldviews*

Every specialized practice system has its principled practice frameworks. These frameworks reflect the specialized practical principled worldview of those working in the principled practice specialty.

Principled practice frameworks exist today for every principled specialty from agriculture to zoning. Many of these specialized practice frameworks have been developing for decades. Professional libraries are full of these frameworks.

Most of the best examples of practice frameworks are found in the principled public professions. These highly principled professional frameworks are frequently challenged by highly novel practice problems. Consider these three very unlikely public professional pronouncements:

- doctor: your malady is beyond the bounds of medical science
- lawyer: your controversy cannot be resolved by legal means
- accountant: your business cannot be valued by accepted accounting methods

All these pronouncements seem unlikely. But why? The reason is that each of these three professions supports a comprehensive, constructive principled practice framework upon which professional practice is based.

It is often said, for example, that "the law is a seamless web." This maxim asserts that the established principles and precedents of the law are sufficient to resolve any legal controversy that might arise.

In this sense these legal principles and precedents of the law can be said to saturate the spectrum of potential legal controversies. So the principles of the law are sufficient to construct new practice in any case where precedent practice is insufficient.

Now you might well live out your days before your condition is cured, your case is settled, or your business is valued. Applicable principle is one thing, applicable practice is quite another.

Master practitioners in all professions often find it necessary to apply available principles to extend the range of professional practice. This process is seldom expeditious or economical.

Every now and again a new principle may be required. High appellate judges, for example, occasionally find it necessary to invent new principles of law. The requirement for new principles generally arises as the result of new practice areas. The Internet, for example, has given rise to the new principles and practices of Internet law.

Sometimes new principles start with new phenomena. The legal phenomenon of the right to privacy is new in $20^{th}$ century U.S. jurisprudence. New pathogen phenomena, such as HIV, emerge in medicine all the time. The economic phenomenon of negative commercial interest rates, of the sort experienced recently in Japan, was once thought to be impossible.

## *Managerial Working Worldviews*

The rise of global cutthroat capitalism has driven radical evolution of managerial worldviews in recent decades. This is capitalist competition taken to a whole new level. Country club management is out. Combat management is in.

The great empires of the past were military empires. The great empires of emerging world capitalist civilization are mercantile and manufacturing and money empires. These new empires compete by means of commerce and credit rather than combat. This new sort of competition is bloodless but no less brutal than that of the old empires.

The top managers of these emerging capitalist empires are the new masters of the universe. Today these elite general managers are rewarded with unprecedented wealth and power. Many are granted perquisites and prerogatives that were once reserved for royalty.

Why are institutional management roles suddenly so much more important in this new cutthroat capitalist world? The reason is managerial superintelligence. Maintaining and maximizing managerial superintelligence is the key to competitive survival and success in the emerging world capitalist civilization.

For present purposes it's useful to distinguish two kinds of managerial superintelligence roles:

- genre manager: institutional administrators of managerial superintelligence systems
- general manager: institutional architects of managerial superintelligence systems

Today elite general managers architect institutional and industry managerial superintelligence systems. This is a highly specialized skill. It requires a very unusual mix of taste, talent, temperament, and training.

Superintelligence architecture requires a working worldview capable of compassing the extreme complexity of diverse, dynamic institutional structures. This working worldview must include all of the key factors related to maximizing management premium.

The modern general management working worldview is not in any sense uniform. Some general managers focus on finance. Others focus on markets and margins. Still others focus on production and personnel.

The working worldview of every general manager must evolve over time. It must change over time to suit changes in institutional challenges and capabilities.

Leadership skills are essential to the success of every general manager. Many great managers have written much about managerial leadership in the popular management press.

Learning to model and manage large-scale institutional system of labyrinthine complexity is at least as important as great leadership today. A great general manager today must be able to cope with multitopical, massively complex institutional system organic order.

Things were very different just a hundred years ago. Until the 20th century management was mostly simplistic and managers were scarce. Most folks lived mainly by the sweat of their brows.

In the year 1900 white collar workers made up about fifteen percent of the U.S. workforce. Today white collar workers make up more than sixty percent of the workforce. All of these white collar workers play some part in mass modeling and mass management.

In the roughest terms the evolution of institutional management can be modeled as:

- agricultural revolution: management as a system
- industrial revolution: management as a skill set
- industrial modernism revolution: management as a mechanistic science

Management, in the modern intuitional leadership sense, is relatively new both as a word and as an idea. Until the 19th century the term "management" in English referred to the routine caretaking of husbandry stock.

Prior to the 19th century most institutions in the western world were limited and localized. Large secular institutions were mostly military such as the Prussian Army and the U.K. Royal Navy. Large sacred institutions included the Catholic Church and various national protestant sects.

These large institutions took the form of specialized monarchies. They employed leadership hierarchies that were essentially nobility structures. In many cases these hierarchies were staffed by the scions of the hereditary nobility. This was a common career path for spare sons and such.

In western civilization the true nobility does battle, dictates policy, and dispenses justice. Nobility does not labor. All forms of active management are considered labor. This includes all forms of oversight, stewardship, supervision, administration, and such.

So management has historically been déclassé in the western world. It has always been something that servants did. In classical antiquity it was mostly something that slaves did.

For this reason management plays very little part in the history of elite western ideas. Great western minds have seldom been engaged by management issues. There is no *Aristotle's Administration*.

Consider China by way of contrast. The arts of administration have a very long history in China. In China a gentlemen is a scholar. A scholar knows the Chinese classics. These classics have much to say about practical administration.

## *Automating Management in the Working World*

Management automation? Can management be automated? Isn't management automation some science fiction dream of a far-off future?

In fact mechanistic management automation is a widespread practical reality today. Mechanistic management automation has long been commonplace in cars, computers, cameras, and cell-phones.

Most of this management automation is out of sight, hidden under the covers. Even so, these managerial mechanisms are surprisingly sophisticated and utterly indispensable.

Computer operating systems represent the state-of-the-art in managerial automation today. This brain-stem level of management automation is hidden under the covers of every kind of computer. The management mechanisms involved are usually invisible unless something goes wrong.

From a managerial standpoint digital computers are automated factories in a box. From the hardware perspective computers are automated bit factories. From the software perspective computers are automated model factories.

These computational factories involve extreme mechanistic complexity. Today the typical personal computer is constructed from many billions of individual parts. Multiple levels of layered mechanistic management automation are required to get all these parts to work in concert.

The mechanics of computational production are very much like those of all other kinds of production. There is making and moving. There is measurement and mandating. There is consumption, conversion, and completion. There are stocks, stores, slates, schedules, and sites.

The computer operating system manages all these mundane mechanisms. Each sort of mundane mechanism is supported by specialized managerial mechanisms. These automated managerial mechanisms, in turn, are managed by automated metamanagerial mechanisms.

Many hundreds of millions of lines of handcrafted code are required to model a state-of-the-art computer operating system today. The clever folks that craft this code don't think explicitly in terms of managerial mechanics. They don't think of what they do in terms of management automation. Even so, this is exactly what computer operating systems are all about.

The myriad mechanisms of computer management are highly specialized. So you cannot use a computer operating system to run an automobile, a bank, or a cruse ship.

Yet very much the same sort of mechanisms are involved in running a computer, an automobile, a bank, and a cruise ship. Very minor respecializations of these mechanisms is sufficient to manage the mechanics of any practical pursuit.

Someday specialized management automation will be routine in many if not most high-value practical pursuits. The Internet data web will serve as a backbone providing connectivity among and permitting collaboration among managers and management automation tools.

This is not some science fiction vision. Rather, it's a technological revolution that is already well underway. Digital computer and communication technologies are already moving inexorably in this direction. This movement is being driven by surging economic forces that have already reached substantial proportions.

Note that this is not yet another utopian artificial intelligence vision. Instead it's an alternative, and mostly antithetical, vision of applied science automation. It is about commoditizing methods of state-of-the-art applied science automation employed in a relatively few high-value applications today.

These state-of-the-art methods are commercially mature and ready for commoditization. The next step is mass custom production of myriad new specialized varieties of mechanistic managerial management automation.

Today institutional management is something that people do, not something that computers do. Computers serve merely as tools in practical management. This is not likely to change in the foreseeable future.

Computers have long been used as advantageous tools in practical management. Most professional managers in the modern industrial world use computers routinely in the course of management.

Even so, computers have vast untapped potential as practical management tools. This is the potential of management meaning automation. Most of the myriad benefits of management meaning automation involve radical reductions of wasted mental motion in management modeling.

Management is always model-intensive. Today computers are good tools for automating managerial model markup media but not managerial model meaning. Managers mostly use computers to prepare documents and to peruse data. This is managerial model media automation.

Managerial model meaning automation operates directly on the mechanistic meaning of management models. The most basic functions of managerial model meaning automation are:

- capture
- check
- coach
- compare
- complete
- compose
- contrast
- correct
- council
- critique

Some more advanced functions of managerial model meaning automation involve:

- scrutiny
- search
- sharing
- simplification
- simulation
- solving
- subscription
- supervision
- surveillance
- survey
- syndication
- synthesis

This sort of managerial machine intelligence is hardly new. Rather it's just the sort of practical language and practical logic machine intelligence that we've been doing on an industrial scale since the 1930s.

The only difference is end-user mass customization. It is exactly this sort of mass custom production that will get us to a radically bigger, better, cheaper, faster version of just the sort of machine intelligence we do today.

Why do we need better management automation? Because all managers always need the best models they can get.

Every manager should have full real time visibility and control of everything everywhere that is germane to that manager's goals. Live strategic, tactical, and operational models of plans and performance should be available at the touch of a button. The ever changing landscape of management objectives, obstacles, and opportunities should be mapped out in excruciating detail.

What will all this look like? It will look and work a lot like the best video games look today. The human brain has a lot more visual bandwidth than verbal bandwidth.

It is no coincidence that all this sounds a lot like the best current examples of military theater and battlefield automation. In the modern world of cutthroat capitalism business, more than ever, is indistinguishable from war.

Management automation is always about supporting managers rather than supplanting managers. It is ultimately about a new optimal division of labor in management.

In this new division of labor computers automate the gears of management. Think of this as managerial robotics.

Gears automation leaves working managers free to focus on the genius of management. Orders of magnitude improvements in managerial productivity and possibility will surely result from this new division of managerial labor.

Will computers ever provide the means to automate the genius of management? Is Jack-Welsh-in-a-box the next step in management automaton?

It turns out that the gears of management are much easier to automate than most think today. It also turns out that the genius of management, for the foreseeable future at least, is impractical to automate. The reason that we can't automate the genius of management is simply that we don't yet have good mechanistic models of what this genius is and how it works.

The genius of management seems to be highly emergent. The genius of management seems to involve large-scale mechanisms that emerge from a diversity of local-scale mechanisms. It is these local-scale mechanisms that we understand as the gears of management today.

Someday we may come to understand the emergence of managerial genius. At this point it may become practical to automate at least some genius of management. Clearly, this point is a long way off in the future.

## *Wiring the Working World Together on the Data Web*

Today the Internet data web should be the ultimate capitalist tool. It should be the worldwide backbone of capitalist civilization.

Someday the Internet data web will be this and more, but not anytime soon. The big problems involve information ideology.

There is only one information technology powerful enough to build out the Internet data web. That one information technology is practical prose. Yet so far practical prose is not part of the plan.

Today the Internet document web and digital media web are both flourishing. Yet the Internet data web, after ten years, is still floundering.

For the past ten years lots of very capable people with considerable cash have labored to make the data web work. These efforts involved communities both in real life and in the research labs. All to no avail so far.

The problem is the vision of the Internet data web as a distributed data processing web. Here the notion is that the data web will grow to incorporate and integrate the vast diversity of institutional data processing installations worldwide. Whole industries will interact seamlessly by means of a world wide web of distributed data processing.

Yet every IT manager understands the severe economic and engineering limitations of distributed data processing. The time, trouble, toil, and treasure involved in even the simplest sort of distributed data processing project today is extreme. So distributed data processing has always been limited to indispensable or inevitable applications.

The limitations of distributed data processing are inherent and insuperable. These are built in from the foundations up. So no reasonable extension or elaboration of distributed data processing practice can hope to overcome these limitations.

The primary problems here are the mentalware and metaphysical bases of universal modeling. Today all data processing practice is based on software science universal modeling techniques. Extracting the practical application mechanics implicit in mentalware and metaphysics is often impractical on small scales. On the massive scale of any interesting Internet data web this is simply inconceivable.

So any sizeable distributed data processing Internet data web involves a massive reengineering catastrophe. There just aren't enough people on this planet to begin to do all the recurrent reengineering required.

If the Internet data web isn't a distributed data processing web, then what is it? It's a worldwide collaborative mass modeling and mass management web. This is a far more practical and profitable sort of Internet data web.

In this approach the Internet data web becomes the ultimate practical informatics platform. The data web becomes the home of all of the practical infodynamics flows of modern mass society. The entire diversity of practical modeling resources will be hosted on the Internet data web.

Today we can mechanize the meaning of less than one percent of the volume and value of management models in service. It will likely take us until the end of this century to get close to one hundred percent.

Why should we mechanize the meaning of all these models? The answer is management productivity and possibility. Maximizing these maximizes management premium.

Economists don't like the idea of management premium because it's difficult to model in purely economic terms. Even so, it's exactly this surplus that is the basis for the unprecedented comforts and contentment of life in modern mass civilization.

We live well because we manage well. We manage well because we model well. Thus we are better able to exploit available mundane resources to extreme advantage. We are better able to extract liquid value from the latent value of these mundane resources.

It is the mass management and mass modeling infrastructure of the modern industrial world that provides unprecedented managerial surplus value. Most of this surplus results from institutional superintelligence.

This is exactly why we must build out the Internet data web. We must do this in order to maximize managerial superintelligence. We must do this in order to maximize management premium.

We must do this because we live in a world where more than two billion people still survive on less than a dollar a day. Mere redistribution of wealth will not solve this or other pressing worldwide problems. Only radical improvements in managerial superintelligence and management premium can, must, and will solve these problems.

A mass modeling and mass management Internet data web requires plug and play information interoperability. So universal modeling is out of the question. End-user utilitarian modeling is the only feasible alternative.

This is how manageable information safety and quality must be woven into the fabric of the Internet data web. This will necessarily require common controlled context practice modeling frameworks.

# 3: The General Theory of Practical Systems

## *Introduction*

GTPS is a systematic approach to understanding the common context of utilitarian modeling. This is the sort of modeling that we all do every day. This is the sort of modeling that managers, professionals, technicians and specialists do for a living.

The intended audience for GTPS is informatics end-users. It is those with elite empirical expository modeling skills. It is intended for those with very high civilization literacy. No computer literacy is required.

The General Theory of Practical Systems is enabling for end-user informatics. GTPS can be viewed as:

- a survey of the common consensus everyday working worldview
- a way of understanding the role of practical information dynamics in capitalist communities
- a foundation framework for end-user Internet data web informatics
- a set of 24,000 primitives for constructing Internet data web structured information models
- a first step towards making computers civilization literate
- an exegesis of the exposition structure of everyday practical English
- a way of understanding mass managerial superintelligence structures
- a systematic way to understand one of the few things that makes us uniquely human
- a map of the landscape and layout of practical logic and practical language
- a synopsis of the common context of practical pursuits and practical reality
- a means of understanding the mechanistic meaning of practical prose and practical programs
- a way of understanding communication and collaboration in the working world
- an overview of the high mechanics of practical modeling and practical management
- a worldly, working education for Internet data web interconnected computers
- enabling technology for mass custom production of Internet data web informatics software
- a modeling mechanism for Internet data web mass management automation

GTPS is not intended for the mainstream information technology audience. The GTPS information ideology is practical information as civilization in disguise.

For this reason those with mainly a universal modeling background may well find GTPS incomprehensible at first. This includes nearly all information technology professionals working today.

From a universal modeling standpoint GTPS may even seem nonsensical. It may seem like nothing more than an extensive exercise in profoundly bad poetry.

From a mainstream information technology perspective GTPS asks and answers all the wrong questions. Moreover, GTPS seems to deny many of the most cherished doctrines both of universal modeling and of mainstream information technology.

All of this is unavoidable. There are excellent reasons for all of this. These reasons involve GTPS aims, architecture, audience, and applications.

## *What GTPS Is and Isn't*

GTPS isn't:

- a universal, unified, ultimate, ultraformalist, utopian theory of everything
- a universal reasoning and representation framework
- a common sense knowledge modeling framework
- a semantic lexicon, semantic frame system, or any sort of linguistics research tool

GTPS is:

- a core set of micromodular mechanistic building blocks for modeling practical meaning
- a foundation framework for high information quality Internet data web informatics
- an approximation of the expository foundations of everyday practical prose
- a framework for mass custom production of end-user Internet informatics

## *The Architecture of GTPS*

The top-level structure of the GTPS framework is:

- foundation
    - language foundation
        - proseology
        - phraseology
    - logic foundation
        - applied mechanics
        - applied mathematics
        - automated modeling
- fundamentals
    - enterprise fundamentals
    - economic fundamentals
    - engineering fundamentals

The bulk of this paper is an overview of GTPS proseology. This is not the entire proseology, just a selection of the high points.

The reason for this emphasis is that the proseology is the starting point for understanding GTPS. Everything else in GTPS elaborates this proseology.

The GTPS proseology is divided into four different levels of reference models:

- GTPS Atlas: major foundation reference models
- GTPS Encyclopedia: minor foundation reference models
- GTPS Differential Dictionary: additional detail differentiation models
- GTPS Topics: specialized differentiation models

Most of the basic exposition structure of practical prose models can be constructed from GTPS Atlas reference models. These are the indispensable invariant structures of the everyday working worldview. They do most of the heavy lifting in routine practice modeling.

The GTPS Encyclopedia and GTPS differential dictionary serves to further elaborate the GTPS Atlas models. These more specialized models provide additional selectivity, specificity, and saturation.

## *GTPS Stenographic Modeling*

GTPS stenographic modeling is the first step toward end-user informatics on the Internet data web. Stenographic modeling provides a quick and easy way to get started with Internet data web GTPS modeling. It provides a convenient way to approach the pencil and paper limit of practical prose meaning in mass collaborative data web modeling.

Stenographic modeling is characterized by:

- structure: free, flexible, and fixed constructive modeling formats
- shorthand: terse, telegraphic, trenchant modeling constructs
- standardization: controlled communal collaboration context standards
- saturation: high selectivity and specificity over the practice meaning spectrum

Stenographic modeling features:

- easy end-user content and context modeling
- simple outline-structured modeling formats
- familiar everyday and elite practice modeling contexts

The killer applications of stenographic modeling include:

- high information quality and safety Internet data web model sharing
- high selectivity and specificity Internet data web searching
- messy multitopical Internet data web modeling mashups

Stenographic modeling structured markup may be:

- visible: directly accessible by means of end-user modeling tools
- veiled: hidden under the covers of Internet informatics applications

The key advantages of stenographic modeling are:

- information quality: up to the pencil and paper meaning limit
- information safety: greatly improved interoperability, insurability, immortality
- simplicity: easy to learn and use for anyone with good exposition skills

Why is formatted modeling necessary? Why is controlled common context modeling necessary? Why not just use all the practical prose sitting on Internet document servers today?
The problem is that unformatted practical prose in the wild, for the foreseeable future at least, is far too difficult to work with. Specifically:

- there are far too many practical contexts with far too much complexity
- practical prose is mixed with personal, political, and other kinds of prose
- the vast variety, variability, and volatility of practical prose styles is truly terrifying

For these reasons and others the prose of mails, messages, and memos is just too hard of a nut to crack[14]. Thus for the next decade or so stenographic modeling will remain the best alternative for data web mass modeling and mass management applications.

## *The GTPS Working Word Processor*

The GTPS Working Word Processor (WWP) is one sort of initial tool for GTPS stenographic modeling. This is essentially a personal computer word processor with a wired-in working worldview.

The WWP is much like any other word processor except that it only works with GTPS working words. These working words are woven into outline-structured practice models. These outline models encode the GTPS working world reference models and GTPS working word differential structures.

The GTPS foundation framework is the starting point for WWP modeling. Specialized modeling communities can add proseology and phraseology models in order to support specialized working worldviews. Providers of topical and target GTPS extensions might include any combination of captive, commercial, and communal sources.

Initially the WWP need only support manual modeling at the language layer of GTPS. This version provides a convenient way of getting started with high information quality modeling on the Internet data web. The applications of this version will include high precision Internet data web structured language serving, sharing, and searching.

This version of the WWP provides a starting point for the evolution of Internet data web authoring and audience tools. In this way WWP can serve the same purpose as the earliest text mode Internet document web browsers (e.g. Lynx).

Further evolution of the WWP provides GTPS logic level support. This will enable WWP model meaning automation features. Applications of this future WWP version will include Internet data web structured logic serving, sharing, and searching.

## *GTPS Methodology Generally*

Practical prose is the ultimate high technology trophy hack. The simple reason for this is that the complexity of the practical prose technostructure is staggering. So how can we tease out the technostructure of practical prose without trivializing it?

In the case of GTPS the goal is not a complete exegesis of the practical prose technostructure. The goal is merely a workable synopsis of the everyday working worldview. The models presented here are a first small step toward that goal.

The redundancy of exposition elements in practical English is extreme. Some, but not nearly all, of this redundancy must be preserved in order to retain most of the power and precision of practical English.

It has long been observed that practical English has a very elaborate alliteration structure. This alliteration structure appears to be accidental for the most part. It's mainly spurious rather than significant.

GTPS exploits this alliteration structure in two ways. First this structure is employed as a filter. Available alliterations provide a convenient way to arbitrarily decide how exposition elements should be selected and structured.

The second way that GTPS exploits this structure is as a tool for improving intellectual ergonomics. GTPS is a big framework. So alliteration is exploited as a tool for maximizing mnemonic and mimetic value. It is exploited as a tool for minimizing mental motion wastage.

Admittedly, some find all the alliterations, adages, and aphorisms profoundly irritating. Yet the only other alternative approach is "word soup." In practical applications it's likely that arranging exposition elements in terms of differential alliteration structures will be a significant human factors advantage.

Consider the term "free". This term appears several times in GTPS. For example:

- free as a metric of variability: fixed, flexible, free
- free as a metric of economic valuation: free, fee
- free as a metric of mobility: fettered, free

All of these senses of "free" are supported by the WordNet[15] framework (wordnet.princeton.edu). Differentiating these senses in a "word soup" framework like WordNet requires looking up the word and searching for the correct sense. In GTPS the differentiation of topical usage is immediately obvious from differential structure.

Now consider the term "flounder". Two of the most common meaning senses of this term are:

- flounder, float
- flounder, fish

These alternative GTPS differential structures provide a rapid way to distinguish these very different meaning senses.

Is the GTPS methodology biased? Yes, but probably not significantly biased. The sense spectrum of everyday practical English seems to be adequately represented.

Note that GTPS is not in any sense an attempt to produce a simplified practical English. Instead the goal is to capture as much of the extreme sophistication of practical English as possible.

So there are lots of ten dollar words in GTPS. These are the words that we use in order to achieve high specificity and selectivity of meaning in our practical pursuits.

There are, admittedly, a few minor coinages in GTPS. In some cases working word meaning senses are stretched to the limit.

This isn't really a big deal. No apologies are forthcoming for these liberties. Just chalk them up to the immunities of prosaic license.

## *Is GTPS Really Meaningful?*

Some people can visualize the mechanistic meaning structure of GTPS right from the start. Most folks with extensive empirical modeling experience need to tinker with GTPS a bit in order to see the meaning structure. Those without extensive empirical expository modeling experience will likely find visualizing the GTPS structure difficult.

Visualizing the GTPS mechanistic invariant structure seems to be particularly difficult for those with an information technology background. It's just not the sort of modeling framework that these folks are skilled at using.

These folks are skilled at modeling with programming language frameworks rather than with practice language frameworks. They are paid for their encoding skills rather than their exposition skills.

GTPS, at least as presented here, provides no axioms or algorithms or any other sort of formalistic modeling. So how can GTPS be anything more than some sort of weird word game? Perhaps GTPS is nothing more than some novel system of sophistry or scholasticism.

The primary goal of the GTPS proseology is sorting out the differential structure of physical and practical reality. Examples include:

- degree: cruel, callous, cold, cool, considerate, concerned, compassionate
- dichotomy: foreign, familiar
- differentia: collective, college, collusion, colony, combine, commune, company
- disjunctions: past, present, prospective
- dimensionality: person, place, period, practice

These simple differential structures express pure practical mechanistic meaning. They are the starting point for more sophisticated differential structures. Differential structures are necessary and sufficient to express the entirely of practical mechanistic meaning in any and all practical pursuits.

Simple differentiation structures serve as models of limited local order. Larger scale physical and practical order is always constructible from layered levels of local order. This is how the differential phenomena structure of physical reality works. It is also how the differential principle structures of practical reality work.

So these differential mechanistic meaning structures are the starting point for sorting out the differential state dynamics structure of practical reality. They are a starting point, but only a starting point, for answering the question "What's the difference?"

Those in the universal modeling world may well be uncomfortable with this approach. The separation of exposition and encoding will surely seem utterly unnatural. The separation of mechanics and mathematics will surely seem unnatural. The separation of fundamental meaning and formal meaning will surely seem unnatural.

Remember that these sorts of separations are what we did in practical informatics before universal modeling. From the 1930s thought the 1970s separate software engineering was the rule. This is what we did before the simultaneous software engineering we do today.

Separate software engineering evolved from manual clerical modeling and early punch card practice. This was a much simpler soft of software engineering. The basic elements were forms, formats, facts, fields, fills, and flowcharts.

These were the days of analysts and coders. Analysts worried about subject matter side informatics modeling. Coders worried about software side informatics modeling.

This division of labor worked pretty well until the 1970s. This is when informatics software complexity rose rapidly as the new 32-bit scalable mainframe architectures advanced. Moreover, minicomputers were beginning to become commonplace.

This is when the separate software engineering separation of labor started to stress. The clash of the very different analyst and coder cultures had always been a problem. Diseconomies of this division of labor grew rapidly with increasing code complexity.

So the modern software engineer was born. This new super-specialist would subsume both the coder and analyst roles. These specialists would see practical reality from a new sort of subject matter and software side stereo perspective. They would do universal modeling rather than utilitarian modeling. They would do simultaneous modeling rather than separate modeling.

Today it's clear that simultaneous software engineering has never been more than just a stopgap. It's what we did to perfect punch card pattern programming and pattern configurable software packages. This is how we followed the path of least economic resistance to where we are now. Today this path gone as far as it's going to go in terms of strategic progress. This is a dead end.

So simultaneous software engineering has run its course. In end-user informatics we must abandon simultaneous modeling for separate modeling. We must return to separate software engineering. There just simply isn't any other choice.

In the age of end-user informatics it's end-users that will do the subject matter side modeling. These folks are the new separate software engineering analysts. Swarms of mass custom software synthesis robots will produce the programs. These are the new separate software engineering coders.

In many ways the end-user informatics revolution is a back-to-the-future revolution. The old school now becomes the newest new school. Replacing simultaneous modeling with separate modeling is just one example of old school practice reborn as on-the-leading-edge practice.

The important thing to remember is that GTPS is not a universal modeling framework. It's a utilitarian modeling framework. This is the sort of modeling that informatics end-users do every day.

In end-user informatics we must always start with the practical logics that informatics end-users understand. They primarily use fundamental mechanistic logics rather than formalist mathematical logics. Mathematics is always an adjunct to mechanics. There are no formalist logics hidden under the covers somewhere. There simply is no need for these.

Aren't these end-users missing a huge opportunity? Shouldn't they just abandon fundamental modeling for formalist modeling? Aren't they missing out on vast modeling and monetary benefits?

In fact elite practice modeling communities are extremely aggressive adopters of beneficial innovations. They don't need much coaxing to change their ways. They rapidly adopt, adapt, and apply beneficial practice modeling innovations as these emerge.

Some in these communities understand formalist modeling well enough to evaluate the potential benefits of various formalist alternatives. These alternatives find few applications in these communities simply because they provide few benefits. This is why universal modeling missionary work in practice modeling communities consistently fails. Universal modeling just isn't competitive with utilitarian modeling in most practical pursuits.

In end-user informatics the end-users are always right. Utilitarian modeling is what end-users do. GTPS doesn't need formalist models because end-users don't need them.

Note that these end-users, as a matter of law, cannot systematically confuse mathematical formalisms and mechanistic fundamentals. They cannot, as a matter of law, do the sort universal modeling commonplace in software engineering.

Especially in the principled professions the choice of modeling mechanisms and mathematics are entirely separable issues. Professional responsibility law standards require the separate justification of mechanisms, mathematics, and the means by which these are combined. Defensive design frameworks demand that these justifications always be documented.

Formalization of practice models always starts with differential mechanistic meaning. This is necessarily true because that's all the mechanistic meaning there is to work with.

There are usually multiple alternative ways to formalize any given range of differential mechanistic meaning. Differential equation systems can be restated as integral equation systems. Algorithms can be restated as axiom systems.

In practical pursuits applicable formalisms serve to encode the differential meaning of practice models. This differential meaning, in and of itself, is always formalism independent.

Moreover, specific formalizations often simplify available mechanistic meaning. Ordinary differential equations may be chosen when partial differential equations are an option. Propositional logics may be chosen when predicate logics are an option. Arithmetic may be chosen when algebra is an option.

For all these reasons GTPS modeling is formalism-neutral. This allows flexibility in choosing formalizations appropriate to any given informatics application.

## *GTPS Context Mining Methods*

Where does GTPS come from? Primarily GTPS comes from fieldwork. It comes from carefully digging through myriad examples of specialized practical prose. This is manual context mining.

This sort of systematic context mining is both an acquired skill and an acquired taste. Very strong practical exposition skills are a prerequisite to doing deep context mining.

Context mining is a skill that is rather difficult to explain to those who have not yet mastered the rudiments. Like computer programming, it's best learned by digging in and doing it.

Those with good practical exposition skills can generally master rudimentarily context mining skills with a little practice. A good culinary cookbook is a good place to start if you'd like to try you hand at everyday practical English context mining. A typical household kitchen provides a pretty good microcosm of practical reality.

Context mining is basically the systematic application of Ockkam's razor. It's about boiling practical reality down to its elementary and elaborate essentials.

Remember that in context mining we're looking at the micromechanical logic element systems of exposition, not the language elements. Mappings from language elements to logic elements are arbitrary.

Context mapping exposes the context structure of practical reality as captured by diverse practical prose samples. GTPS context mining prose samples are selected from the professional practice literature of engineering, economics, law, medicine, finance, management, accounting, and many varieties of specialized enterprise. So the everyday practical English of the principled professions is the primary source for GTPS.

On the whole the sense spectrum of this slightly elite sort of practical English prose seems much more familiar than foreign. It doesn't seem to be that much different from the sense spectrum of other sorts of everyday practical English prose. There are just a lot more ten dollar words.

The place to start in context mining is finding the top-level exposition elements. The key to top-level context mining is fishing out the indispensable mechanistic meaning invariants. These are the main mechanisms one must always use in practical prose modeling.

Which sorts of micromechanical meaning invariants are utterly indispensable in practical pursuits? Which exposition invariants, when removed, make modeling practice impossible?

Some of the most obvious examples of indispensable mechanisms in practice modeling are person, property, place, period, practice, and policy. That's why these important exposition elements are included in the GTPS primary primitive set.

Any significant sample of practical prose involves multiple parallel, pluralistic perspectives. Comparative proseology is the way to sort these out. This technique examines sets or spaces of exposition invariants. Some major examples include mechanics, supply, and control.

This technique involves selectively removing or replacing expository element structures in practical prose samples. The differential meaning that results from these changes exposes the differential structure of the underlying context space.

Exposition element changes impact the practical meaning of a prose sample in various ways. Some mechanistic meaning invariants have much more impact than others.

Sometimes removing meaning elements renders the prose sample meaningless. Other times the meaning of the prose sample is changed or degraded. Many meaning elements do not impact the practical mechanistic meaning at all. These meaning elements are personal, poetic, pious, and so on. These are purely subjective for practical purposes.

Note that GTPS supports a wide variety of subjective word senses. This is necessary in order to support the bulk of routine practical communication. We seldom write pure practical prose in the course of routine practical pursuits. If you try this sometime you will find that it's remarkably difficult.

GTPS supports language modeling of subjective meaning, but not logic modeling. We simply don't know how to model or mechanize this sort of meaning. So GTPS logic tools, given what we know today, must simply skip over subjective meaning elements.

Topical context mining cannot be performed in a vacuum. Some considerable topical subject matter mastery is required to do elite prose context mining. We all have most of the modeling mastery required when it comes to everyday practical pursuits. This is not the case in specialized practical pursuits.

Elite principled practical pursuits require a systematic approach to context mining. Here artifacts are just as important as articulations. So two complementary kinds of context mining are required. These are:

- context archeology: working with topical mundane and managerial artifacts
- context anthropology: working with topical modeling standards, stocks, staffs, and systems

Both are required because it's necessary to examine both things and models of things simultaneously. This is necessary in order to understand how working word constructs map to working world mechanistic meaning.

Systematic manual context mining techniques of this sort have long been employed in some specialized sorts of automation science fieldwork. Adaptations of these techniques are the starting point for exposition engineering in end-user informatics.

## *Mechanized Context Mining Methods*

GTPS is the product of proprietary manual methods of manual context mining. Mechanized context mining is an alternative approach[16].

It is hardly secret that major Internet search providers have an intense interest in mechanized context mining technology. Novel methods of mechanized context mining are a key factor in many recent Internet search engine technology innovations.

Some see automated methods of computerized context mining as a way to produce very high quality context frameworks for various applications. Much work is being done in this area both in academia and in the IT industry.

GTPS is the product of extensive manual context digging and detective work. Can mechanized context mining produce an alternative to GTPS? If so, how will these mechanistically mined context models compare with manual context mined models like GTPS?

Mechanized context mining methods involve various forms of mathematical pattern recognition. Today the most promising of these methods involve nonlinear multivariate statistical pattern inference.

Methods of multivariate statistical inference and induction[17], as in the case of Bayesian techniques, work on the language of practical prose. It can't work directly on the mechanistic logic of practical prose samples because this meaning is implicit.

In commonplace practical prose the mappings between language patterns and logic principles is astonishingly complex. There is not just one of these mapping but untold millions of these mappings. Moreover, practical meaning maps are always confused with personal, philosophical, and other sorts of meaning maps that we don't even begin to understand.

This is why modeling statistical language patterns without supporting practice logic principles can be downright dangerous. There can be no automatic way to model information safety and quality. Finding interesting patterns is never enough.

There is seldom any shortage of statistical patterns. Some of these patterns are significant while others are spurious. In most cases there is no reliable means to sort these out automatically.

Moreover, you can always find the statistical pattern you're looking for if you try hard enough. Any economic time series can be explained entirely by the phases of the moon with very minor sorts of statistical sleight of hand.

Note that all forms of pattern recognition always involve some degree of sample space structural simplification. This is very often radical simplification. Regularity is filtered in and randomness is filtered out.

The goal in producing high quality context frameworks is to capture as much of the sophistication of context structures as possible. This requires preserving randomness and referential richness. Statistical and other forms of pattern recognition in general are not well suited to preserving these aspects of the invariant structure of complex sample spaces.

All of this severely limits the potential power of mechanized context mapping. Mechanized context mining produces pattern context frameworks of language encoding. These simplistic context frameworks may well be highly advantageous for various Internet document web search engine applications.

Internet document web search engines don't worry much about information quality or safety. Document web search results always include a lot of noise and nonsense. Better pattern models can help to reduce the noise and nonsense.

End-user Internet data web informatics is a whole different story. Here information safety and quality are paramount. There is no room for noise and nonsense. Context logic and language must be carefully controlled. Small context errors will generally result in substantial communication and collaboration errors. So Internet data web informatics foundation frameworks must be rock solid.

Mechanical principles are the most powerful way we know of to express practical meaning. Getting these principles right is prerequisite to manageable information safety and quality. Pattern practice modeling, so far, provides a radical simplification of the best-practice principled modeling at best.

Thus it seems unlikely that mechanized context mining will compete with GTPS in high information quality informatics applications. For the foreseeable future statistical pattern mined contexts will not be adequate for end-user informatics, informatics industrialization, or mass management automation.

The goal of GTPS is to provide context models sufficient to support the pencil and paper practical meaning limit in end-user informatics applications. For the foreseeable future this sort of high information quality informatics will start with context frameworks produced by manual context mining methods.

This does not mean that there is no place for pattern modeling in high information quality informatics. There may very well be substantial synergy between mechanized and manual methods of context mining.

Moreover, is seems likely that at least some GTPS applications will employ hybrid methods combining a mix of pattern modeling and principled modeling. This may be the optimal approach where an application can tolerate a mix of high and low quality informatics models.

## *GTPS Model Theory*

The only GTPS global model is present practical reality itself. There is no top level, global GTPS theory of everything (TOE).

All GTPS models are micromodular. Some are more general than others. Some are considerably larger than others. Reference model primitive counts range from two up to several hundred.

There is a set of GTPS primary primitives that provide a very general way to parse the meaning of practical reality. These are merely provided a convenience. They are hardly canonical-closed. They may or may not be useful for any particular GTPS application.

OK, then how does one do general purpose reasoning? How does one prove the global correctness, completeness, and consistency of GTPS itself and of GTPS applications?

The simple answer is that you don't. There is no such thing in this universe as general purpose mechanistic reasoning, apart from the trivial varieties. Other universes can fend for themselves.

General purpose practical reasoning is a GTPS non-goal. Instead the goal of GTPS is to support mass custom production of specialized forms of topical and target genuine reasoning.

GTPS does not build in any formalist system of type theory, model theory, or proof theory. Any appropriate modeling formalism may be applied to GTPS models. All such formalizations are always adjuncts to the GTPS architecture.

Every practical informatics application requires an appropriate balance of encoding formalisms and exposition fundamentals. Note, however, that encoding and exposition are always entirely separable issues.

GTPS enforces separation of these issues by means of encoding neutrality. Thus the fundamentals of the GTPS foundation are suitable for any sort of appropriate encoding.

## *GTPS Logic Layers*

Ultimately GTPS must serve as a starting point for mass custom production of applicable algorithms. There is, of course, no universal way to do this. We know this because there is no universal way to do mechanistic representation and reasoning in this universe. There are just lots of useful and often unique ways.

So there is no such thing as a universal practice modeler. There is no such thing as a universal software synthesizer.

Common sense readily confirms this. There is no such thing as a universal computer program. There is no such thing as a universal sales report. There is no such thing as a universal hand tool.

Thus the only feasible approach to informatics mass production is mass custom production. We must mass custom produce a great diversity of specialized software synthesizers. We must mass custom produce a great diversity of specialized practice modeling tools.

The GTPS logic layers support practice modeling semantics in terms of three different sorts of complementary semantics spaces. These include:

- symbolic algebra: applied mathematics semantics
- state automata: applied mechanics semantics
- symbol automata: application modeling semantics

Representation and reasoning on these logics is not in any sense general. It is always specific to well defined, highly disciplined specialized semantics spaces. These are the topic and target modeling spaces combined in the context of some specialized practical pursuit.

The semiotics of GTPS modeling involves mappings within and among these various spaces. These mappings are how mechanical, mathematical, and model automation meaning is combined in the GTPS logic framework.

Needless to say these are incredibly messy mappings. These mappings will be mostly local to targets and topics. It is these mappings that are always the advent of algorithms in practical informatics.

Where do applicable algorithms come from? Applicable algorithms are always accessories. They merely serve to automate and animate applied mechanics models and applied mathematics models.

Remember that computation, per se, has no applied mathematics or applied mechanics meaning. Computation in isolation merely has metalogical meaning. So applicable computation gets its meaning by mapping analogically to some combination of:

- applicable mechanics: applied state automata
- applicable mathematics: applied symbolic algebras

There is, of course, a great diversity of each of these. We can't even begin to guess how many there are or could be. So far there is no systematic way to map these diversities in their entirety.

There are, however, lots and lots of well known and widely used cases. Synthetic software mass custom production methods will mature over time by tackling manageable ranges of special cases. This will require lots of descriptive information science work.

The vast majority of applied mathematics applications today can be and are done on spreadsheets. Higher mathematics applications, like finite element analysis and convex programming, are very high value but relatively low volume applied mathematics applications.

Much of the landscape of applied mathematics and applied mathematical algorithms has been pretty well mapped out at a very high level. This is a pretty messy map, but it works well enough for routine applications.

Mathematical rapid application development (RAD) tools implement much of this map. MATLAB from MathWorks (mathworks.com) is an excellent example of this sort of tool.

MATLAB supports a wide variety of applied mathematics models. These include primitive models such as matrix operations and analytic equation solvers. In addition MATLAB supports numerous specialized application frameworks. Some examples include robust control, digital signal processing, digital image processing, and financial mathematics.

MATLAB models are widely used to synthesize both hardware and software for various embedded control applications. Your cell phone may very well contain hardware and software of this sort.

Applied mechanics is another matter. So far we really don't know very much about the large-scale landscape of applied mechanics. Hopefully GTPS will prove a useful starting point for exploring this landscape.

One result of this exploration will likely be RAD software tools for applied mechanics. This sort of tool includes micromodules supporting all the usual mechanisms of daily living. Add-on micromodule packages might include home automation, investment portfolio management, medical condition management, and so on.

Note that there is a classic canonical-closed way to model applicable mechanisms. This is general automata theory. This is another example of an old school engineering science model which is largely neglected today.

It is easy to demonstrate that every applicable mechanism can be decomposed into state dynamics automata structures. A state dynamics automata is a mechanism that can be characterized adequately in terms of interface and internal state dynamics.

Here again, there is no general way of classifying state dynamics automata. There are a few dozen major kinds and a few hundred minor kinds that find frequent application.

How do we get from mechanistic automata to model automation algorithms? The way to get started is brute force.

One way to look at an algorithm is as a symbol dynamics automata. The simplest way to synthesize applied mechanics algorithms is to naively map symbol dynamics spaces directly to state dynamics spaces. Programmers do this all the time. Automation of this sort of mapping is employed in certain very specialized automation science applications.

Note that once again there is no general way to do this. Mappings between symbol dynamics and state dynamics are always arbitrary, analogical, and anecdotal. These mapping are workable only over well defined, disciplined ranges of local order.

In commercial algorithm mass custom production far more sophisticated methods will generally be employed. These are very much the same sort of methods that find various applications today in various parts of the automation software world.

## *GTPS vs. CYC*

Today there are dozens of universal modeling foundation frameworks in the research world. These employ various sorts of pattern representation and reasoning methods.

The most ambitious of these frameworks are known as commonsense knowledge bases. The primary focus of these artificial intelligence frameworks is natural language recognition, representation, and reasoning research.

The current state of the art in commonsense knowledge bases is unquestionably CYC[18] from Cycorp of Austin, Texas (cycorp.com). CYC is short for cyclopedia but also a pun on psyche.

CYC began as a pioneering "hard" artificial intelligence research project. Doug Lenat and the CYC team have been working to perfect CYC since 1984. Today CYC is finding application to Internet Semantic Web context modeling applications.

One of the original aims of CYC was to construct an autonomous artificial intelligence capable of representing and reasoning the entire meaning of an academic encyclopedia.

Today CYC is still a very long way from this very ambitious goal. Yet all agree that CYC is a brilliant achievement.

How do GTPS and CYC compare? In many ways these are almost exactly antithetical. They are radically different in nearly every respect. The questions asked and answered by CYC and by GTPS are almost entirely contrary. So one way to look at GTPS is as an anti-CYC.

CYC employs a proprietary artificial intelligence language (CYCL) and operates on families of ontological frameworks. GTPS is implementation-independent and employs mechanistic rather than metaphysical frameworks for modeling meaning. These are micromodular differential state dynamics frameworks.

CYC and GTPS start with nearly antithetical legacies. CYC builds of the research legacy of the "hard" artificial intelligence communities. GTPS builds on the real world legacy of the automation software and automation science communities.

Perhaps the most important difference of all is that CYC aspires to model the meaning of every sort of human language and logic. This includes personal, poetic, philosophical, pious, and so on.

The focus of GTPS is strictly limited to mechanistic practical meaning. GTPS supports subjective language only as a convenient means of decorative detailing.

GTPS makes no attempt to support subjective logic. GTPS logic modeling is limited to differential state dynamics logics. These are the mechanistic Newtonian logics of enterprise, economic, and engineering practice modeling. They are the logics of mechanistic managerial modeling.

GTPS has nothing to do with autonomous individual artificial intelligence. Instead, the emphasis is on automating the institutional intelligence of mass modeling and mass management.

This is mechanistic managerial machine intelligence. This is the sort of practical machine intelligence that is the basis for all practical informatics today. This is the sort of machine intelligence that has been widely employed in diverse practical pursuits for the better part of a century.

CYC is ultimately about perfecting mental machine intelligence. It's about understanding individual cognitive intelligence. It is about the dream of creating computers in our own image. It's about automating and animating universal representation and reasoning. It's about unlocking the putative power of universal language and universal logic.

Perhaps someday CYC will achieve all these goals. This is not likely to happen any time soon. So far CYC has found only a very limited range of advantageous applications.

## *GTPS in Context Maintenance Automation*

Practical context maintenance is a huge and highly important social dynamic in modern industrial civilization. This is the process that maintains the perfection of topical and target practice modeling contexts. These diverse contexts are the foundation upon which all modern mass modeling and mass management systems are built.

Today context maintenance is mainly the responsibility of the elite of the principled practice modeling elite. This is something like ten percent of the three hundred million people that do elite principled practice modeling as some part of their working lives.

For present purposes assume that this involves thirty million people worldwide. Furthermore, assume that the economic effort involved has a value of about three hundred billion dollars annually.

Much of this activity has already become established on the Internet document web. The next step getting topical and target context maintenance moved onto the Internet data web.

GTPS is a first step in moving practical context maintenance onto the Internet data web. This is the first step in the evolution of end-user Internet data web informatics.

Note that no manageable community of people can possibly succeed at doing context maintenance. So we must rely on just those utterly unmanageable communities that do context maintenance today. We must convince these communities to move their efforts to the Internet data web.

How is context maintenance done? It's a dynamic dialectic decision process. It's an ongoing argument. Authors and audiences endlessly argue everything out.

All these topically and target specialized context arguments needs to be moved onto the Internet data web. Some consensus authoring mechanism much like Wikipedia (wikipedia.org) will be required to host these communities. Experiments of this sort underway today are known as "semantic wikis[19]."

So far few efforts to support topical and target contexts on the Internet data web have succeeded. This is true both for the schemaware and semantic data web architectures.

The problem is universal modeling. Software and schemaware informatics standards are universal modeling standards. These are pattern modeling standards.

Standards of this sort, as information technology veterans well know, are far from definitive. Moreover both first principles and practical experience confirm that these standards cannot reasonably be made definitive.

The reason that such standards cannot be definitive is lack of explicit mechanistic meaning. Pattern informatics standards are modeled in terms of metaphysics and mentalware. So the implicit subject matter mechanics of these standards is always open to wide interpretation.

Explicit mechanistic practical meaning must be recovered from pattern standards by means of reengineering. The problem is that everyone gets a different answer from this reengineering. Not surprisingly this is always exactly the answer they need.

The result is rapid fragmentation of pattern informatics standards. One valid *de jure* standard becomes many variant *de facto* standards. Reference implementations, implementation notes, and lots of tinkering are required to make each variant *de facto* version of the standard work.

All this is remarkably expensive. Moreover, it seldom works at all for non-trivial informatics models. Most successful non-proprietary software and schemaware informatics standards are relatively simple.

Successful informatics standards, even the simplest ones, often come with a large entourage. There folks are standards consultants and contractors. They have test suites and test beds. They have reference implementations and regression tests. They have compliance testing and certification procedures.

Those that use the informatics standard ultimately pay to support this entourage. Informatics standards compliance costs in some cases can run to many millions of dollars. So software and schemaware standards compliance can create barriers to entry and efficient competition in the informatics world.

Other practical pursuits employ and endless variety of definitive practice standards. These are principled rather than pattern practice standards. They have explicit mechanistic subject matter meaning. The automotive and aerospace industries have many thousands of these standards.

It has been said that standards are so important that everyone wants their own. Any definitive standard must saturate its intended practice mechanics spectrum with sufficient selectivity, specificity.

Principled informatics standards, done well, can provide this sort of definitive standardization. These standards must necessarily employ utilitarian rather than universal modeling methods.

GTPS utilitarian modeling enables this sort of definitive principled informatics standardization. Standards of this sort are a prerequisite to moving the social dynamics of practical context maintenance to the Internet data web.

### *GTPS in Pervasive Internet Automation*

Pervasive Internet computation is the ultimate Internet computing platform. Pervasive Internet automation is the killer application for this platform.

The pervasive Internet computation revolution is already well underway. Computers of all kinds are turning up everywhere in everything. Today computers are commonplace in automobiles and appliances, toys and tools, plus an endless variety of indispensable digital gadgets.

Today there are hundreds of computers in you life. Soon there will be thousands. By the middle of next decade there will likely be a trillion computers in service worldwide. Most of these will be embedded, under the covers, computers.

Disposable printed electronics paper and plastic computers are the next stage in pervasive computation. Packaged foods, clothing, wallpaper, books, newspapers, and the like will soon come with Internet accessible computers as standard equipment.

Pervasive Internet automation is about combining all these disparate Internet computers into networks that can automate the dynamics of modern mass civilization. The payoff comes from the synergy value of combining disparate local automation into large-scale automation.

Combinations of this sort will always be highly specialized and most will be unique. GTPS enables these mass custom pervasive automation combinations both as glue and as grease.

As glue GTPS provides the common context for pervasive automation mashup modeling. As grease GTPS provides an easy end-user way for everybody to model how all the disparate computers in pervasive automation must work in concert.

## *GTPS in End-User Internet Informatics*

End-user informatics is do-it-yourself, demand-driven, disposable informatics. It's about having just the informatics solutions you need just when you need them.

The infrastructure and instrumentation costs of end-user informatics will be quite substantial but the incremental cost will be quite small. This gives a whole new twist to the notion of free software.

The roles of end-users in end-user informatics include:

- target: design new informatics applications
- tailor: modify existing informatics applications
- test: using automated test case generation and validation

None of this involves any sort of computer literacy whatsoever. This is not some sort of programming in disguise scheme. The entire informatics lifecycle is modeled and managed entirely in the context of end-user working worldview language and logic contexts.

In end-user informatics everybody in every practical pursuit automatically becomes a software author. Mass custom end-user informatics software will be produced automatically as a byproduct of the routine managerial modeling we must always do anyway.

The most important sort of informatics end-users are managers, professionals, technicians, and specialists. These folks will produce the lion's share of informatics software in the coming age of end-user informatics.

These talented folks are the expository modeling elite. They are the principled practice modeling elite. They are the institutional intelligentsia of modern mass society.

It is these folks do nearly all of the high value principled practice modeling. They are the starting point for the synergistic emergence of institutional superintelligence.

These folks are always civilization literate but seldom computer literate. This is as it should be. The fashionable notion in software science academia that these folks must all eventually become computer literate is simply nuts. Computational thinking[20] is never a suitable substitute for civilized thinking.

End-user informatics starts by making computers civilization literate. So there is no need to make civilization computer literate.

It is the expository modeling end-user elite that will do most of the work required to make computers civilization literate. These folks are the only reasonable choice to do this work. They are the civilization literacy authorities, authors, audiences, and arbiters.

How will they do this? They with gradually teach their computers about how the world works over time.

They will start with the everyday working worldview foundation. They will educate their computers by elaborating and extending this foundation. They will add their own specialized working worldview models to the mix. In this way they will fill in and flesh out all of the topic and target contexts they use.

GTPS is a starting point for civilizing our computers. Even so, GTPS alone provides only the most elementary of educations. Diverse specialized practice communities will need to furnish the educational extensions that they need.

This will be an automation science rather than an artificial intelligence education. GTPS is an expository framework foundation for massively multitopical for automation science mechanistic meaning modeling.

Some in the automation science world will be skeptical of this claim with good reason. This is a well known hard problem in automation science. This is a Holy Grail problem in automation science. Nearly all experts in automation science consider this to be an intractable design problem if not an impossible one.

How does GTPS solve this thorny technological problem? The answer is that GTPS cheats. GTPS changes the question. GTPS is not a designed foundation framework. GTPS is a dug up foundation framework.

Why does end-user informatics need an expository foundation framework? The answer is that a solid practice modeling foundation is critically enabling technology. An adequate expository foundation is indispensable in order to solve the leverage and linkage problems intrinsic to massively multitopical mechanistic practice modeling.

The leverage problem is about eliminating micromodular redundancy across diverse topics and targets. The linkage problem is about doing deep and disciplined micromodular referential binding across diverse topics and targets.

There is really only one workable solution to this problem. This is the obvious and optimal solution.

Just use the expository foundations of everyday practical prose. Just use the same foundation framework that informatics end-users already understand. Just use the same solution that already works every day everywhere in the modern working world for all practical purposes.

Any other solution is just plain crazy. No manageable group of people is every going to succeed in designing an alternative solution. Even if they somehow miraculously succeed their alternative would be utterly, completely, totally redundant.

Moreover, all of our familiar end-user practice models would have to be laboriously restated into this second, utterly foreign, context. This is encryption modeling rather than exposition modeling.

GTPS avoids all this madness. GTPS is an exegesis of the mass consensus expository foundations of practical prose. So GTPS is an ideal starting point for providing end-user informatics leverage and linkage.

End-user Internet informatics is about enabling elite principled practice modelers to move their mass modeling and mass management systems onto the Internet data web. This begins by hosting the socially-intensive process of topical and target community context maintenance on the Internet data web.

Why would these communities move these processes onto the Internet data web? The answer is major reductions in wasted mental motion.

In all practical pursuits mentation is a specialized, scarce, and steeply priced resource. Today our use of this resource is all too often spendthrift.

The tactical benefits of model meaning automation always involve reducing wasted mental motion. In a perfectly automated world no modeling decision would ever be made more than once in all of history.

Good practical modeling is always expensive and effortful. Typically ninety percent or more of practical modeling effort involves wasted mental motion. Much routine practical modeling involves ninety nine percent or more wasted mental motion.

This high redundancy in manual practice modeling can be radically reduced by means of model meaning automation. Today there are many hundreds of commercial and captive varieties of automation science model meaning automation.

We seldom use this very specialized sort of software but frequently benefit from it. This is the sort of software that designs automobiles, balances mutual fund portfolios, and models new pharmaceutical manufacturing processes.

The most sophisticated sorts of this software are often found in automated modeling factories today. In these factories a single model automation seat typically costs tens or hundreds of thousands of dollars per year. This is just the total technology cost. These are tool chain and intellectual property costs. Someone to sit in the seat, management costs, and other overhead costs are extra.

Where can we find the smartest software around today? The usual answer is IBM's Deep Blue chess playing software. This is the one that beat world chess champion Garry Kasparov a few times.

This software is smart in the "soft" artificial intelligence sense. It uses sophisticated optimization algorithms to map out the extreme combinatorial complexity of chess move pattern spaces.

For all practical purposes Deep Blue is just a stunt. It's a really cool stunt, but a stunt nonetheless. It's a toy rather than a tool.

Note that Deep Blue isn't nearly the smartest software at IBM. IBM has many of the best electronic and mechanical automated modeling factories around.

These are serious tools rather than toys. These are capitalist weapon systems for competing in the cutthroat global computer hardware market.

Tools of this sort are easily the smartest sort of practical software found anywhere today. This is the sort that does state-of-the-art practical meaning automation.

So far this sort of practical meaning automation modeling is always narrowly monotopical. It lets us say almost anything about almost nothing. Universal modeling, on the other hand, allows us to say almost nothing about almost anything.

So neither of these alternatives supports the entire spectrum of practical meaning today. Only practical prose supports this entire spectrum today.

Practical prose lets us say as much as needs to be said about anything practical. We can do this because practical prose has solid expository foundations. This foundation provides all of the multitopical leverage and linkage that anybody ever needs. It is just because of these foundations that practical prose lets us say as much as we wish about anything we wish to say.

GTPS is an automation science approximation of this foundation. So GTPS supports roughly the same foundation spectrum of practical meaning as the foundations of everyday practical prose.

In this way GTPS enables the extension of existing mechanistic methods of model meaning automation from monotopical to massively multitopical modeling. This is how we can, must, and eventually will routinely automate practical model meaning up to the pencil and paper limit.

Can we automate the exposition structure of practical prose? We're already pretty good at this. Note that the model structure of the best mechanistic meaning automation models is indistinguishable from that of practical prose. This is not surprising. Both kinds of modeling must accurately reflect the messy, mechanistic minutiae of practical reality.

So automation science model meaning automation is really just utilitarian modeling in disguise. It is monotopical practical prose language and logic automation in disguise.

The next step in automation science is extending model meaning automation to massive multitopicality. GTPS is indispensable enabling technology for this next step.

This next step is the starting point for end-user informatics. End-user informatics is about commoditizing, commercializing, and commoditizing messy, managerial automation science model meaning automation. It's about turning the Internet data web into the automated model factory for the entirely of modern mass civilization.

## *GTPS in Informatics Industrialization*

Massively multitopical meaning automation is a mass custom production proposition. The reason is that there is no such thing as general purpose utilitarian model automation. We know this conclusively from practical experience. We know this with certainty from examining the principle structure of practical reality and from the phenomena structure of physical reality.

What do these structures look like? They are essentially random. They can be visualized by looking up at the sky on a starry night.

The order we see in the night sky is sparse. There are diverse kinds of local order, hints of regularity, and a few major landmarks. There is a whole lot of randomness.

This is just the same kind of order we see when we drive down the street or fly overhead in an aircraft. This is just the same kind of order we see when we look thought a telescope or microscope. This kind of order involves diverse sorts of irregular recurrence at all scales swimming in a sea of randomness. Reality is mostly random independent of range or region.

So end-user informatics is necessarily about mass custom production of genuine purpose utilitarian model automation. It's about supporting the specific, specialized random genuineness of real life practice.

Each of these specialized sorts of genuineness is multitopical. So end-user informatics must support massive multitopicality. We can only do this by mass custom production support for monotopicality.

We must mass custom produce monotopical practice modelers and software synthesizers. We must use the GTPS as a foundation mashing monotopical models up into messy, multitopical managerial models.

We must mass custom produce automated end-user informatics factories. We must build automated modeling factories to mass custom produce these modeling factories.

This is the only way we can achieve model meaning mechanization up to the pencil and paper limit. This is the only way that we can automate and animate the full flexibility, fidelity, and fusion of practical prose modeling.

Where will we find the models for industrializing end-user informatics? These won't be the pattern models we use today in information technology universal modeling[21]. They must be the principled practice models that informatics end-users already have.

Yet this is not the direction in mainstream information technology today. Here the notion of model-driven[22] software engineering is currently fashionable. Model-driven architectures, model-driven development, model-driven engineering, and such receive a lot of coverage in the software trade and technical literature.

The notion here is that universal modeling can be perfected by adding multiple layers of metamodel patterns. The implication here is that if your patterns aren't powerful enough then adding more patterns is the solution.

This simplistic approach has often been tried before. The result is always the same. Augmenting patterns with more patterns rapidly degrades information quality and safety. Both first principles and extensive practical experience confirm that pattern model meaning is not accretive.

The real problem with patterns is patterns. The one great virtue of pattern modeling is generality. The well known vices of patterns include very limited safety, scalability, specificity, and selectivity.

This is why very high standards of mechanistic principled modeling are mandated and maintained in nearly all practical pursuits today. This is why pattern modeling is both obsolete and outlawed in nearly all principled practice.

The only models that really count in practical informatics are end-user models. These are the ones that practical people routinely use in their practical pursuits. These are principled models rather than pattern models. These are mechanistic models rather than metaphysical models.

Can we dispense with pattern modeling in practical informatics? We must, at least in end-user informatics. This may be a blessing in disguise. The truth is that pattern modeling has never made a lot of sense in practical informatics

Note that pattern restatements of principled practice models have always been utterly redundant in the first place. So every dime spent on these restatements is money wasted. Plus there's the problem of all the smart parts of these models getting systematically stripped away in the process.

So it makes perfect sense to just use exactly the practice models that our end-users already have as a basis for practical informatics industrialization. These, conveniently, are already perfect for all practical purposes.

The obvious way to do this is end-user application modeling. No more craftwork punch card pattern programming. No more pesky programmers.

Can we get rid of informatics programmers? Yes, we can, and sooner or later we must. Handcrafted pattern programming just isn't ever going to get us to informatics industrialization.

Heaven knows that there are much better things for software science folks to do than sit in cubicles crafting very much the same code day after day. In the era of end-user informatics these folks will work use automated tools to mass custom produce applicable algorithms.

These mass custom algorithms will provide the basis for automated software encoding. Automated encoding of this sort is mostly a solved problem today. This is entirely manageable with minor adaptations of current commercial technology.

Note that automated software encoding is not in any sense automation of handcrafted pattern programming. That is simply impossible. Automating any form of pattern programming involves multiple major miracles both of people and of physics.

Automated encoding is merely about mechanistically mapping state dynamics automata to symbol dynamics automata. This can be accomplished by automation science means that range from the very simple to the highly sophisticated.

Automated encoding produces disposable software from durable subject matter models. It produces informatics middleware from institutional intelligence modelware.

Automated encoding is informatics infrastructure independent. Synthetic encoding tool cluster components will support multiplatform synthesis.

A synthesis target platform is some convenient combination of informatics infrastructure. A target platform combines infrastructure choices such as operating systems, programming languages, software foundation libraries, communication stacks, and the like.

End-user informatics synthetic software production is intellectual property-intensive mass custom software production. This will is capital-intensive mass custom software production.

End-user informatics software synthesis will resemble the ways in which digital hardware has been synthesized to great advantage for over twenty years. Hardware is just frozen software, after all.

The one big bottleneck problem in mass custom informatics modeling is feedstock. Synthetic encoding starts with empirical exposition models suitable for encoding. Today we don't have this mechanistic practice model feedstock.

GTPS is enabling for portable end-user informatics modelware. GTPS will engender commercial informatics modelware IP markets. Other sources of informatics modelware IP will include captive and communal providers.

Synthetic mass custom informatics production will be done in automated informatics factories. These factories will be similar to the sorts of automated model factories that have long been commonplace in industries such as aerospace and automotive. The difference is that automated informatics factories will operate collaboratively worldwide on the Internet data web.

Information engineering specialties in the age of end user-user informatics will include:

- infodynamics engineers: mass modeling and mass management engineering science
- informatics engineers: software and subject matter framework engineering science

Informatics engineers will specialize software-side and subject matter-side informatics frameworks for specialized audiences. Varieties of informatics framework specializations in the age of end user-user informatics will include:

- international informatics framework specialization
- industry informatics framework specialization
- institutional informatics framework specialization
- individual informatics framework specialization

Informatics engineering specialties in the end-user informatics age will include:

- encoding engineering: computer literate software side informatics engineers
- exposition engineering: civilization literate subject matter informatics engineers

Most current software professionals will eventually migrate to one of the informatics engineering specialties. Both will work to maintain topical and target specialization of informatics factory tool chains and modelware stocks.

Primary exposition engineering roles include:

- sages: deep specialized subject matter modeling expertise
- scribes: exposition framework modeling and maintenance expertise

The vast majority of informatics engineers will work in scribe roles. This is a technician role requiring a two year junior college associate degree. These folks will maintain institutional frameworks and assist end-users in building and using their specialized informatics tools.
In this way end-user informatics will replace high skill, high salary labor with far lower skill and lower salary labor. This is the sort of labor for capital tradeoff typical of mature capital-intensive mass production.

GTPS modeling is intended for use in systematic methods of exposition engineering. It is these methods that will provide the feedstock for synthetic software encoding.

Note that this is a much more balanced approach to informatics than that which prevails today. There are no longer any elite empirical exposition specialists in best current software engineering practice.

In end-user informatics a new, better balance of encoding and exposition engineering will reflect the range of modeling done in particular end-user informatics communities.

Mass production is not magic production. Mass custom production starts with mastery of customized mechanics.

Informatics end-users already have all this mechanistic mastery. This mastery comes in endless specialized varieties. Informatics engineers will work with end-users to automate informatics modeling of all these specialized varieties.

Informatics engineers will work with end-user context communities to automate and animate topical and target mechanistic context modeling spaces. These engineers will specialize in those varieties native to the settings in which they work. They will mass custom produce end-user informatics tools for specialized topic and target audiences.

End-user communities will use these specialized tools to build Internet data web informatics applications. The architecture of these end-user applications will be massively micromodular. These highly customized informatics architectures will mirror and mimic the messy mechanistic minutia structure of practical reality.

# 4: A Quick Tour of GTPS Highlights

## *Introduction*

The dynamics of practical reality are emergent. The complexity of this emergence is enormous. This is why practical prose has evolved such a great diversity of alterative practical perspectives. This diversity of perspectives is how we divide and conquer the complexity of practical reality.

Each perspective provides a specialized set of micromechanical reference models suitable for sorting out some aspect of practical reality. Perspectives are combined opportunistically to suit the circumstances at hand.

The following is a synopsis of some of some of the more interesting and important GTPS perspectives in the GTPS reference model atlas. In most applications these perspectives, employed in various combinations, will capture most practice model meaning.

## *Primary Primitives*

There are about a hundred primary primitives in GTPS. These are a top level decomposition of physical and practical reality. The most useful of these include:

- person: an individual or institution
- property: an investment or interest
- place: a point, patch, or pocket in space
- period: a instant or interval in time
- practice: making, moving, managing, etc.
- policy: prescriptions, prohibitions, etc.
- principle: an emergent enterprise, economic, or engineering mechanism
- phenomena: an elementary physical or practical mechanism

## *Control*

Control is an example of highly reusable rhetorical mechanism in practical prose. This is not surprising since management is complex conative control. The mechanics of control permeates most commonplace sorts of practice modeling.

Working words like can, will, must, should, may, are words that express the language and logic of practical control. In addition there are a great many very specialized varieties of control language and logic.

No one thing, in and of itself, is very interesting for any practical purpose. The dynamics of practical pursuits are combinatorial.

The dynamics of nails needs the dynamics of hammers. The dynamics of getting need the dynamics of giving. The dynamics of shopping needs the dynamics of selling.

The simplest form of control is dual differential dynamics control. This involves minimizing the difference between to sets of dynamics:

- conation: the mandated dynamics of wishing, willing, wanting
- cognizance: the measured dynamics of the world as it is, was, or will become

The primary models of combinatorial dynamics are:

- coevolution: combinatorial change
- coequilibrium: combinatorial condition
- coelasticity: the potential for combinatorial equilibrium

Coping is the most general model of control in practical pursuits. Coping is about:

- care: anything of value
- concern: connation vs. cognizance
- capability: the possibility of conduct
- capital: things the provide capability
- contrivance: a choice of a conduct plan
- conduct: change in response to concern

The primary varieties of communal control are:

- collaboration: the compatible dynamics of allies
- coordination: neither collaboration nor competition
- competition: the contrary dynamics of adversaries

Leadership generally involves a mix of:

- command: leadership by authority
- consensus: leadership by agreement

## *Artifact, Asset, and Administration*

Everything in practical reality is an artifact. An artifact is any product of human arts. Artifacts have:

- authors: persons that design the artifact
- audiences: persons that use the artifact
- applications: ways that an artifact can be used

Most artifacts have many applications.

- actor: any artifact organizational role
- agent: any proactive artifact organizational role
- article: any passive artifact organizational role
- action: any artifact role
- actuation: any proactive artifact operational role
- activation: any passive artifact operational role

Administration is the mechanics of asset management. The primary elements of administration are:

- administrator: any manager of assets
- administration: any combination of administrators
- answerability: accountability relationships among administrators

The elements of answerability are:

- advice: providing measurements
- assent: permitted mandates

Every administrator role involves:

- authority: the right to authorize asset uses and action utilization
- accountability: responsibility to account for authorize uses and action utilization

Every administrator is responsible for some range of affairs. Specifically:

- affair: any required asset actualization
- agenda: any combination of affairs

## *Factorization*

Factorization provides a canonical-closed framework for modeling systematic order in terms of forms and functions. Specifically:

- factor: any form or function
- form: any factor of system organization
- function: any factor of system operation

A formulation consists of:

- formation: any combination of forms
- functionality: any combination of functions

## *Fitness*

Fitness is the mechanics of freedom from flaws and faults. The primary elements of fitness are:

- 
- fitness: functionality suitable for some intended use
- failure: any disorder that limits artifact applicability
- flaw: any failure of organizational form
- fault: any failure of operational function
- fix: any means of restoring fitness

Flaws and faults are often related. A broken automobile fan belt is a flaw. The overheated automobile engine resulting from this flaw is a fault.

Varieties of freedom from failure include:

- fine: flawless and faultless
- flawless: absence of non-trivial flaws
- faultless: absence of non-trivial faults

## *Production and Praxeology*

Production theory is the differential dynamics view of getting things done. The key elements of production dynamics are:

- platform: things used in production
- plant: things that produce
- product: things that get produced
- protocol: things undertaken in production
- procedure: production efforts
- process: production effects

Praxeology is a comprehensive managerial view of production. The key elements of praxeology are:

- purpose: production problems and provocations
- pragmatics: models necessary for production planning
- planning: determining what to produce
- pursuit: doing production
- perfection: production

## *Orchestration*

Orchestration is the mechanics of operational optimization. Orchestration involves choosing sequences of states that maximizes operational objectives. The primary varieties of operational states are:

- obligatory state: a prescribed operational state
- optional state: a permitted operational state
- obstacle state: a prohibited operational state
- outlaw state: a proscribed operational state

## *Mechanics*

Mechanics is the dynamics of mortal and mineral mechanism. Primary varieties of mineral mechanisms include:

- money
- material
- merchandise
- machinery
- model

Mechanisms are characterized in terms of:

- manifestations: instantaneous mechanism state
- modifications: interval mechanism state

Primary varieties of modifications are:

- mundane
    - make
        - materialize
            - maturation (e.g. growing a crop)
            - manufacture (e.g. assembling refrigerators)
        - maintain
            - minister (e.g. water the lawn)
            - meliorate (e.g. add a garage to your home)
            - moderate (e.g. insulate your home)
            - mitigate (e.g. fix windows broken by a storm)
    - move
        - manual movement
        - mechanized movement
- managerial
    - measure (e.g. balance your checkbook)
    - mandate (e.g. order a pizza)

## *Technology*

Technology is the dynamics of tools and techniques:

- tool: anything useful
- technique: any use of a tool

Primary varieties of mundane technique are:

- transformation: make technique
- translocation: move technique

Technology gets us thought our tasks:

- task: any useful undertaking
- target: the intention of a task
- team: people using tools
- treatment: a combinations of techniques
- toolkit: a combination of tools

## *Resource*

Resource is the dynamics of realizations and revisions. Resources are roleplayers in revision routines. Routines are modeled in terms of:

- recipe: requirements and results
- rations: reservations and requisitions
- rules: regulation and regimentation

Varieties of repetitive revisions include:

- rote: repetitive making
- route: repetitive moving

## *Instrumentality*

Instrumentality is the dynamics of instruments and implementations. Combinations of these are involvements. Varieties of involvement include:

- integration: combinations of instruments
- interoperation: combinations of implementations

Primary varieties of integration include:

- interconnection: information integration
- interlink: move integration
- interlock: make integration
- induction: kymatic integration
- impact: kinematic integration

Primary varieties of interoperation include:

- interaction: make interoperation
- interchange: move interoperation
- intercommunication: a sequence of interlocutions and inferences
- influence: kymatic interoperation
- impulse: kinematic interoperation

## *Enterprise*

Enterprise is the economic view of community. Enterprise models the dynamics of systematic extraction and exploitation. The primary varieties of exploitation are:

- extraction: turning latent value into liquid value
- exploitation: any means of extraction

Primary varieties of exploitation include:

- enjoyment: passive exploitation such as appreciation and arbitrage
- employment: active exploitation
    - engagement: using but not using up

- exchange: use in giving and getting
- expenditure: using up

The economic dynamics of an enterprise may be modeled by:

- equity: cumulative earnings
- earnings: exaction less expense
- exaction: economic value resulting from exploitation
- expense: the economic value required for exploitation

## *Supply*

Supply is about the dynamics of person and property flows in production. The elements of supply are:

- supply
    - stock: property supply
    - staff: person supply
- site: supply places
- schedule: supply periods
- slate: supply practices

## *Society*

Society models the dynamics of service. Social networks are service networks. Varieties of personal service include:

- stipulation: symbiosis service
- supervision: submissive service
- suppression: subjection service

Primary varieties of social networks include:

- state: secular society
- sect: sacred society

Primary social networks include:

- social contract: social stipulations
- social conflict: social struggle
- social class: social stratification

## *Management*

Management is the dynamics of complex conative control. The differential dynamics of management involve:

- measure: any mechanistic model of the world as it is, was, or will be
- mandate: any mechanistic model of the world as it could be, can be, or could have been

Managers cope with concerns given available capital and conduct alternatives. Specifically:

- matters: concerns
    - mission: the what of concerns
    - motive: the why of concerns
- means: capital alternatives
- methods: conduct alternatives

## *Interest*

Interest is the dynamics of property utilization. Interests arise both out of investment and indebtedness. The primary varieties of interest are:

- title: full or factional property ownership interest
- temporary: occupancy without ownership interest
- tentative: imperfect or immature interests interest
- trust: possession interests, possibly in perpetuity, by intermediaries and fiduciaries
- trade: interests arising in commerce and other forms of trade

## *Reciprocity*

Reciprocity is the dynamics realization roleplayer relationships. Relationships may be characterized as:

- rivalry: cooperation relationship
- rapport: competition relationship

Neither rivalry nor rapport is permanent.

- rupture: transition from rapport to rivalry
- rapprochement: transition from rivalry to rapport

The primary dynamics of reciprocity are:

- request: asking
- response: answering
- report: advising
- reaction: acting
- reminder: admonition

Resource reliance results in one or more reckonings:

- redemption: roleplay suitable to a role
- reckoning: request for redemption

The outcome of a reckoning may be:

- requite: adequate roleplay
- renege: abstinence of roleplay
- renounce: abnegation of roleplay

## *Deontology*

Deontology is the dynamics of communal dealing and domination:

- dealing: communal coping by consensus
- domination: communal coping by command

Deeds start with decisions. The elements of the decision cycle are:

- deliberation: consideration of candidate conduct
- decisiveness: choosing candidate conduct
- doing: coping by means of conduct

Varieties of dealing include:

- dickering: mutual advantage cooperative dealing
- disputation: mutual animosity competitive dealing

The elements of duty are:

- deliverable: what is to be done
    - dictation: what must be done
    - discretion: what may be done
- deadline: when the deliverable must be done

Domination is modeled in terms of:

- domain: demesne + denizens + dominion
- denizen: domain person
- demesne: domain places
- dominion: domain discipline policies and practices
- discipline: the execution and enforcement of dominion
- domicile: the residence of a domain denizen
- demography: any model of the denizens of a domain

## *Community*

Human civilization is the ultimate community. Larger communities are compositions of smaller communities. Some element of community is:

- culture: the models of a community
- cambistry: the money of a community
- corpus juris: the legal system of a community
- ceremony: community rites and rituals
- centralization: authority vested in chiefs, constables, etc.
- communal values: virtues and vices
- criminality: crime and punishment
- communication: communal information flows
- catering: providing the necessities of life
- caste: communal segregation and stratification
- cornucopia: critters and crops
- companionship: communities of two
- children: the primary thing that civilizations produce
- chrematistics: communal capital flows within a

## *Affluence*

Affluence is the accumulation of assets. The primary elements of affluence are:

- accumulation: any combination of assets
- acquisition: any supplementation of accumulation assets
- alienation: any subtraction of accumulation assets
- assignment: an alienation plus an acquisition of an asset

Primary varieties of assignment include:

- accommodation: intentional assignment
- arrogation: involuntary assignment
- accidence: inadvertent assignment
- alms: indulgence assignment
- abandonment: indifference assignment

## *Change*

Everything is always changing. The dynamics of change over the lifecycle of any realized thing is:

- creation: The first instant of a career
- career: the further dynamics of a career
- cessation: The final instant of a career

A career is a sequence of changes. Varieties of change include:

- condition: a career instant
- chapter: a career interval
- course: a sequence of courses and conditions

Much change is cyclical. The elements of cyclicality are:

- cycle: any recurrent change
- cadence: the repetitiveness of a cycle

Change is explained in terms of:

- causation: engenderment in combinatorial dynamics
- contribution: exchange in combinatorial dynamics
- conducion: enablement in combinatorial dynamics

## *Commonality*

Commonality is combinatorial similarly. The elements of commonality are:

- comparison and contrast
- classification and categorization
- composition and compounding

Primary varieties of combination are:

- composition: combination of components
    - collection: a set ordered composition
    - consecution: a sequentially ordered collection
    - cue: a dynamic consecution
- connectivity
    - construction: a schema ordered composition
    - circuit: a schematic ordered composition
    - continuum: a spatially ordered composition
- compound: combination of constituents

The important difference between this is:

- component: the things in a composition
- constituent: the kind of things in a composition

Thus the components of a pair of shoes are a left shoe and a right shoe. The constituents of café au lait are coffee and milk.

The dynamics of combination are modeled by:

- cumulation: the set dynamics of combination
- choreography: the structural dynamics of combination

Primary varieties of comparison are:

- correspondence: comparability of form
- correlation: comparability of function

Varieties of dynamic correlation include:

- contravariance: contrasting dynamics
- covariance: comparable dynamics

Varieties of correspondence include:

- congruent: comparable shapes that match
- conjugate: contrasting shapes that mate

Calibration is comparison to a calibration standard. Two commonplace kinds of calibration standards are:

- chronograph: any model for calibrating space
- cartograph: any model for calibrating time

## *Existential*

Existence is the totality of extants. Extants have extent and endurance. Extants evolve. Primary varieties of practical evolution are:

- effectuation: enterprise evolution
- exploitation: economic evolution
- energetics: engineering evolution

The dynamics of evolution can be modeled in terms of:

- event: evolution at an instant in time
- episode: evolution over an interval of time

Etiology models the dynamics of evolution in terms of:

- exertion: energy exerted
- evocation: evolution exhibited

# 5: GTPS Primary Primitives

## *Introduction*

The GTPS primary primitives are the foundations of the GTPS foundation. There are just over a hundred of these primitives. A few dozen are sufficient to get started.

At the top level GTPS starts differentiating primary primitives as:

- applied mechanics primitive
    - physical reality primitive
    - practical reality primitive
- applied mathematics primitive
    - arithmetic
    - algebra
    - analysis

These GTPS primary primitives are the basic building blocks of practical reality. Among the most useful of these are:

- person: e.g. you
- property: e.g. you house
- place: e.g. you city
- period: e.g. the current hour
- practice: e.g. cooking lasagna
- policy: e.g. thou shalt not steal
- principle: e.g. drive on the correct side of the road
- phenomena: e.g. traffic conditions for the morning commute

These primary primitives are answers. The questions that these primitives answer are just the usual questions in everyday expository modeling. These questions are:

- who: person
- what: property
- when: period
- where: place
- why: policy
- how: practice, principle, and phenomena

## *Person Practical Reality Primitives*

The notion of personification is important in all forms of practical modeling. Because of personification General Motors and General Grant are indistinguishable for many purposes. General Motors can't vote, but it can pay taxes and defend a lawsuit.

Thus the top level divisions of person are:

- personage: a vital human being
    - individual: a living personage
        - pre-natal individual
        - post-natal individual
    - interee: a lifeless personage
- persona: a virtual human being
    - institution: any human enterprise
        - private institution
            - family private institution

- connubial family: family by matrimony
- consanguine family: family by bloodline
- firm private institution
- pecuniary firm: private firm primarily for profit
- philanthropic firm: private firm not primarily for profit
- public institution
  - secular public institution
  - sacred public institution
- industry: any enterprise of enterprises

You are an individual and hopefully will remain one for some time to come. You are a member of a family, the largest single sort of private institution. You secondary school is most likely a public institution if you were raised the developed world.

The Church of England is an example of a clerical institution in the United Kingdom. This is an established religion and therefore a public institution. The Catholic Diocese of the United Kingdom is, for tax purposes at least, a private philanthropic firm.

A personage engaged in practice is a practitioner. We are all practitioners of something. Some other examples practice roles include:

- pal, playmate, and paramour
- patron, pensioner, and petitioner
- payer and payee
- physician, pharmacist, and patient
- potentate, prince, patrician, and plebian, and peon
- procurer and provider
- professor and pedagogue
- promoter and popularizer
- propinquity, progenitor, parental, and progeny
- proprietor and personnel
- proselytizer and proselyte
- proxy, plenipotentiary, and puppet
- pulpit and parishioner
- pundit and prophet
- pupil, prentice, and protégé

## *Property Practical Reality Primitives*

Property is anything with any utility. Note that the utility value of property is not always in the plus column. A drum of radioactive waste is example of property with substantial negative economic value.

For much of human history the notion of property has been pretty simple. Something like:

- praedial property: land and the things attached to it (e.g. estate)
- personal property: everything else (e.g. effects)

The modern notion of property can get quite complicated. The elements of property are:

- mundane property: property with an identifiable physical existence
  - movable property: e.g. your car
  - motionless property: e.g. your house
- managerial property: property without any identifiable physical existence
  - intangible property: bank accounts, leaseholds, etc.
  - intellectual property: ideas a property e.g. patents, trademarks, etc.

Mundane property is property in the material world. Managerial property is about rights and responsibilities in the conduct of management.

Varieties of moveable property include:

- mortal property
    - living property: crops, livestock, pets, even people in some places
    - lifeless property: hides, furs, meat products, etc.
- mineral property
    - machinery property: cookware, tools, furnaces, etc.
    - material property: raw material, spare parts, fuel, etc.
    - model property: books, statues (model media but not model meaning)
    - merchandise property: saleable property such as cars, clothing, food, etc.
    - monetary property: physical, coins, cash, and equivalents

The notion of movable vs. motionless property goes back to many of the earliest legal systems. Varieties of immovable property include:

- spatial property: property as place
    - earthly property: property on the earth
        - tellurian property: land only property
        - terraqueous property: land + water property as in a lakefront cottage
        - thalassic property: water property such as an ocean reserve
    - extraterrestrial property: property elsewhere
- structural property: immovable improvements
    - edifice property: buildings, barriers, bridges, etc.
    - earthworks property: roads, drainage, dams, etc.
- situated property: sky and subsurface immovable property
    - extractable property: water rights, mineral rights, etc.
    - environmental property: air rights, quietude, sunlight, etc.

If you own a house in the suburbs then you have a piece of tellurian property. In some counties you can only own the structural property (the house itself) but the government owns the spatial property (the land).

In the U.S. you may or may not own the extractable property rights at your home site. So if oil bubbles up in you back yard you may not be in luck after all.

How about Europa, the forth largest moon of the planet Mars? As a matter of law that's public property in most jurisdictions. There are elaborate international treaties about extraterrestrial property rights.

Varieties of intangible property include:

- interest property: any potential benefit from sort of property
- intellectual property: the meaning of a model as property
- interpersonal property: relationships with family, friends, etc.
- indenture property: rights to apprenticeships, bondsmen, etc.
- immunity property: the right to avoid some sort of burden such as criminal prosecution
- indemnity property: the right to recompense such as health insurance
- indulgence property: a right to do something as property as in a currency market option

Property has the potential to produce some sort of benefit. All forms of interest property are useful benefits arising from property investment or indebtedness. If you own an olive grove then you also have the interest in the olive harvests produced by your grove.

You need not own property in order to have an interest in a property. You might arrange to sell the rights to a harvest of your olive grove as a property interest. This interest is independent of the olive grove itself. You keep the grove but must supply the buyer of the harvest interest with the fruits of the grove.

Suppose someone builds an open air music theater next to your nice home in the suburbs. How might this impact your property situation?

You still own your home and can live there. So there is no change in your motionless property situation.

There is, however, an impact on the benefits available from your home. It's no longer a reliably peaceful place to live. Having a peaceful place to live is useful and thus valuable. Therefore you have suffered an intangible property loss. Your interest in the peaceful enjoyment has been impaired. Thus you legitimate interests are economically impaired.

Indenture is involuntary servitude without slavery *per se*. Slaves are persons as property. Indenture of prisoners, children, and others is still practiced today. This is most often true in primitive and pre modern societies.

Whose property are you? You are your own property. Even so, you can't sell yourself in the modern industrial world. Thus you are your own inalienable property. You may have other forms of inalienable property as well such as the right to vote (an indulgence) and the right to privacy.

Your children are not your property but you have an interpersonal property right to their companionship. Damages in wrongful death suits are based on this sort of property.

An important distinction involving property is:

proprietorship: person owning property (property that you have) possession: person occupying property (property that you hold)

So if I borrow you lawnmower then I am its possessor rather than its proprietor. You are the lawnmowers proprietor but not, for the time being, its possessor.

## *Place Practical Reality Primitives*

Primary varieties of place primitives include motionless and movable places. A motionless place is always the same place like that Straits of Gibraltar. A movable place changes over time like the trunk of your automobile.

Varieties of place include:

- motionless place: any static place
    - point place: a one dimensional place like a set of GPS coordinates
    - patch place: a two dimensional place like the boundaries of you garden
    - pocket place: a three dimensional place like the dimensions of a room
- movable place: any shifting place

Place is conveniently combined with period, person and property for place dynamics modeling. Specifically:

- placement
    - parking: property at a place
    - posting: personage at a place
- presence
    - persistence: parking for a period
    - perseverance: posting for a period

Some common varieties of placement include:

- pinpointed
- penned
- packed

- perched
- projecting
- protruding
- penetrating
- pervading
- permeating
- proximate

## *Period Practical Reality Primitives*

Period primitive varieties include:

- instant: one specific point in time
- interval: a span in time specified by zero, one or two instants
    - inveterate interval: an interval with only a specific final instant
    - interim interval: an interval with a specific first and final instants
    - interminate interval: an interval with only a specific first instant
- indeterminate interval: an interval with no specific instants

The year 2000 is an interim interval period because its first and final instants are well known. The lifespan of smallpox as an endemic human disease is an inveterate interval. The final instant of this interval occurred in 1997. The first instant is still a mystery.

Your lifespan is an interminate interval, at least so far. The first instant is known but the final instant lies in the future.

Tense is an important property of periods. Varieties of period tense include:

- past tense
- present tense
- prospective tense
- past present tense
- present prospective tense
- past prospective tense (past + present + prospective)

Any two periods may be compared as:

- preceding: aforetime
- parallel: at the same time
- proceeding: aftertime

Some periods are longer than others. The length of a period may be characterized as:

- precarious: lasting seconds
- passing: lasting minutes
- perishable: lasting days
- pro tempore: lasting weeks or months
- persistent: lasting years
- permanent: lasting many years or decades
- perseverant: lasting may centuries of millennia
- perennial: lasting for eras or eons
- 

Varieties of special periods include

- primordial: since the beginning of time
- perpetual: for all time
- perdurable: until the end of time

## *Practice and Phenomena Practical Reality Primitives*

Practice is the genesis of everything in practical reality. The primary varieties of practice are:

- mundane practice
    - make practice: e.g. baking a chocolate cake
        - materialization practice
            - maturation practice: e.g. raise a child or a crop
            - manufacturing practice: e.g. build a house or a car
        - maintenance practice: e.g. wash the dishes
    - move practice: e.g. getting yourself to work
- managerial practice
    - measure practice: e.g. counting lawnmowers in stock
    - mandate practice: e.g. ordering more lawnmowers for stock

The weather is an important part of practical reality? Who practices the weather? Nobody, of course, at least amongst the human race.

Even so, most natural processes can be considered as principled practice for most practical purposes. This requires a touch of fiction. Specifically it requires the following fictional distinction:

- providential: apparent aleatory purpose practice
- premeditated: actual artificial purpose practice

Practice can always be modeled in terms of applied science mechanisms. These emergent mechanisms are principles. The primary varieties of practice principles are:

- enterprise principle: e.g. stocking, staffing, salary administration, etc.
- economic principle: e.g. accounting, finance, bidding, buying, etc.
- engineering principle: e.g. material handling, manufacturing mechanisms, etc.

Practical principles can be usually be reduced to practical phenomena. Phenomena are modeled in terms of:

- proportionality: correlation among combinations of phenomena
- parameterization: model value spaces

Note that both of these may be qualitative, quantitative, or mixed.

## *Policy Practical Reality Primitives*

The realm of practice is ultimately constrained both by physics and by policy. Physics disallows anti-gravity while policy disallows anti-competitive pricing.

The elements of policy are:

- policy: constraints on the who, what, why, when, and where of practice
- policymaker: one that makes policy

The primary varieties of policy are:

- public policy: e.g. legal systems generally
- private policy: e.g. your company's office supplies procurement policy

Primary varieties of practice policy include:

- sanction: policy do's
    - precept: must always
    - prescription: must
    - permission: may

- suggestion: policy druthers or policy discretion
    - preference: might
    - pococurante: maybe
    - prejudice: might not
- stricture: policy don'ts
    - premonitory: may not
    - prohibition: must not
    - proscription: must never

Where does policy come from? There are many sources of policy. Some of these sources are highly subjective. Some primary sources include:

- code policy
    - enterprise policy: produciblity code
    - economics policy: profitability code
    - engineering policy: possibility code
    - equity policy: permissibility code
- custom policy
    - ethics policy: probity custom
    - etiquette policy: pleasantry custom
    - esthetics policy: prettiness custom
    - eudemony policy: pleasure custom
    - ethnic policy: patrimony custom
- creed policy
    - electoral policy: political creed
    - ecclesiastic policy: piety creed

A permit is a grant of practice permission. Specifically:

- permit: practice permission
- permit holder: one with practice permission

Some varieties of permit include:

- patent: an exclusive practice permission
- pass: a passage permission

## *Physical Primitives*

The starting point for modeling physical reality is physical state. Recall the following concepts your high school textbook:

- physical state: the embedding of physical substance in physical spacetime
- physical substance: any form of matter or energy
- physical spacetime: three dimensions of space plus one of time

Note that physical state always has physical shape. The dynamics of possible shapes are the basis for all physical laws.

So the top level physical primitives are:

- spacetime physical primitive
    - physical space: classical Galilean space in most practical applications
    - physical time: classical Galilean space in most practical applications
- substance physical primitive
    - matter substance
        - corpuscular matter: molecules, atoms, sub atomic particles, etc.
        - condensed matter: solids, liquids, gases, plasmas, etc.

- energy substance
  - motion energy: kinetic (pieces, particles) and kymatic (wave) energy
  - motionless energy: all forms of potential energy such as a cell phone battery
- state physical primitive
  - possibility physical primitive
  - phenomena physical primitive

What sorts of states are possible in physical reality? This starts with three different kinds of possibility criterion:

- physical conservation: physical state dynamics conserve physical substance
- physical conservatism: physical state dynamics are always least action
- physical coherence: physical state dynamics is compact, connected, and continuous

You can't be in two places at once. Things don't appear magically out of nowhere. You can get there from here.

These are all consequences of physical conservation, conservatism, and coherence. All the various sorts of physical mechanics we use start with specialized assumptions based on these fundamental physical constraints.

Note that certain very particular violations of these assumptions are allowed in theoretical physics at very small and very large scales. Quantum mechanical non locality is one example of this sort of violation.

- basic science phenomena
  - material phenomena: physical chemistry, crystallography, etc.
  - motion phenomena: radiation mechanics, solid body mechanics, etc.
- natural science phenomena
  - mortal phenomena: biology, ecology, etc.
  - mineral phenomena: geology, meteorology, etc.
- applied science phenomena
  - production phenomena: the foundation phenomena of enterprise
  - price phenomena: the foundation phenomena of micro and macro economics
  - potential phenomena: the foundation phenomena of engineering

Price theory is presented in every introductory economics course. Production and potential theory are mostly unknown today outside of scattered and highly specialized modeling communities.

Production theory is an elaboration of the work done by Fredrick Winslow Taylor in the early years of the 20$^{th}$ century. Production theory provides a canonical basis for modeling the mechanics of all kinds of mundane and managerial production.

Potential theory is basis for energetics modeling in large scale engineering projects such as oil refineries. The physical mechanics of making and moving can always be reduced to systems of potential theory models.

## *Primitive Pluralities*

Some useful pluralities of primary primitives are:

- portfolio: plurality of property
- personnel: plurality of practitioner
- population: plurality of person
- plat: plurality of place
- project: plurality of practice
- progression: plurality of period
- pandect: plurality of policy
- paraphernalia: plurality of property

- potpourri: plurality of property

# Internet Resources

Related material is available on the Internet at http://end-user-informatics.blogspot.com and at http://www.theultimatetechnology.com.

# Notices

Copyright © 2007 by George A. Maney (gmaney<at>ieee<dot>org). All rights reserved.
CYC and CYCL are registered trademarks of Cycorp (cyc.com)
Deep Blue is a registered trademark of IBM (ibm.com)
Ford is a registered trademark of The Ford Motor Company (ford.com)
IBM is a registered trademark of IBM (ibm.com)
MATLAB is a registered trademark of The MathWorks (mathworks.com)
System/360 and OS/360 are registered trademarks of IBM (ibm.com)
UML is an Object Management Group (omg.org) registered trademark
Universal Modeling Language is an Object Management Group (omg.org) registered trademark
UNIX is a registered trademark of The Open Group (unix.org)
World Wide Web is a registered trademark of the World Wide Web Consortium (w3.org)

For the record, note that all statements made in this book are IMHO.

# About the Author

George A. Maney lives in California's Great Central Valley and works as a management information science consultant. His extensive professional education and experience includes management, law, applied mathematics, automation science, electrical engineering, and computer science.